\documentclass[%
 reprint,
 superscriptaddress,
aps,
amsmath,amssymb,
pre,
]{revtex4-2}

\usepackage{physics}
\usepackage{graphicx}
\usepackage{dcolumn}
\usepackage{bm}
\usepackage{bbold}
\usepackage{algorithm2e}
\SetKwComment{Comment}{/* }{ */}
\usepackage[usenames,dvipsnames]{color}
\usepackage[hidelinks]{hyperref}
\usepackage{mathtools}

\usepackage{yhmath}

\DeclareUnicodeCharacter{2061}{}

\newcommand{\nyuphysics}{Center for Soft Matter Research, Department of Physics, New York University, New York 10003, USA}

\newcommand{\nyusimons}{Simons Center for Computational Physical Chemistry, Department of Chemistry, New York University, New York 10003, USA}
\newcommand{\nyucourant}{Courant Institute of Mathematical Sciences, New York University, New York 10003, USA}
\newcommand{\nyucns}{Center for Neural Science, New York University, New York 10003, USA}

\newcommand{\fashion}{FaSHIoNPOp}

\begin{document}
\preprint{APS/123-QED}

\author{Mathias Casiulis}
\email{mc9287@nyu.edu}
\affiliation{\nyuphysics}
\affiliation{\nyusimons}
\author{Stefano Martiniani}
\email{sm7683@nyu.edu}
\affiliation{\nyuphysics}
\affiliation{\nyusimons}
\affiliation{\nyucourant}
\affiliation{\nyucns}

\date{\today}

\title{Fast generation of spectrally-shaped disorder, on the sphere}

\begin{abstract}
The design of disordered point patterns with desirable properties is an exciting and ongoing research endeavor, with applications ranging from materials to computer science.
A successful approach in recent years has been the optimization of point patterns through a loss function that enforces properties in their Fourier-space representation.
Yet, these methods have so far strictly been limited to flat Euclidean space, precluding their use in the contexts of coatings of curved surfaces for photonics, or of sampling of curved manifolds for instance.
We introduce~\fashion, an algorithm that relies on fast non-uniform spherical harmonics transforms, to enforce pair correlations in point patterns on the sphere with an $O(N \log N)$ complexity in $N$ the number of points.
Having demonstrated its performance, we showcase applications of~\fashion, ranging from the generation of hyperuniform structures on the sphere for sampling and physical applications, to the design of gyromorphs (disordered structures with maximal scattering power at a given frequency) on the sphere.
We additionally show that~\fashion~can be combined to both global constraints like centrosymmetry, with applications to the design of more isotropic $3d$ gyromorphs, and local real-space constraints like pair repulsion.
Our work paves the way for optimal sampling and coating design on curved manifolds, with many applications across physics, materials and computer science.
\end{abstract}

\maketitle

\section{Introduction}

Correlated disorder, the family of random structures that display correlations but lack the translational order of crystals or quasicrystals, have emerged as a new paradigm across materials design~\cite{Florescu2009,Man2013,Leseur2016,Froufe-Perez2017,Bigourdan2019,Haberko2020,Piechulla2021,Monsarrat2022,Klatt2022,Vynck2022,Vynck2023,Casiulis2024a} and sampling~\cite{Yan2015,Pilleboue2015,Pharr2018}.
While a growing family of systems spanning physics~\cite{Torquato2018,Pine2005,Corte2008,Wilken2020,Weijs2015}, chemistry~\cite{Martelli2017}, biology~\cite{Jiao2014,Tarnita2017,Huang2021,Djeghdi2022,Liu2024b}, ecology~\cite{Ge2023}, and social systems~\cite{Dong2023} have been shown to display emergent long-range correlations in disordered structures, many open questions remain on the realizability~\cite{Yamada1961,Kuna2007,LachiezeRey2014,Galerne2015,LachiezeRey2015} and properties~\cite{Vynck2022,Vynck2023,Casiulis2024a} of subclasses of correlated disorder, making numerical generation strategies invaluable, and even useful in practice to manufacture structures~\cite{Florescu2009,Man2013}.

One such approach, the so-called collective variable method~\cite{Uche2004, Uche2006,Morse2023}, relies on the optimization of a loss function to generate point patterns with specified correlations.
While the method traditionally relied on costly algorithms, with scalings as bad as $O(N^3)$ with $N$ the number of points, a recent breakthrough~\cite{Shih2023}, with the Fast Reciprocal-Space Correlator (FReSCo)~\cite{FReSCo}, brought this scaling down to $O(N \log N)$ thanks to the use of non-uniform Fast Fourier transform at key steps of the calculation.
In spite of its wide applicability, FReSCo remains fundamentally limited to Euclidean space, thus precluding the generation of correlated disorder of curved surfaces.
This limitation critically affects applications in sampling, as many problems involve measurements on the surface of spheres (\textit{e.g.} in acoustics~\cite{Rafaely2005,Pinardi2021,Tanaka2025}, electromagnetism~\cite{Marantis2009}, geophysics~\cite{Chao2014}, computational fluid dynamics~\cite{Flyer2012}, atmospheric physics~\cite{Ishioka2018} or rendering~\cite{Yan2015,Pilleboue2015,Pharr2018}), but also applications to photonics, in which coatings of curved surfaces have so far assumed near-flat surfaces~\cite{Vynck2022}, thus limiting the validity of models and the expressivity of design strategies.

In this paper, we propose an optimization algorithm analogous to FReSCo but on the surface of the sphere, the Fast Spherical Harmonics Implementation of Non-Uniform Point-pattern Optimization (\fashion).
Taking advantage of fast multithreaded implementations of non-uniform Spherical Harmonics Transforms (NUSHTs)~\cite{ducc}, we show that~\fashion~ achieves quasilinear computational cost like FReSCo for arbitrary targets.
This work paves the way for large-scale generation of correlated disorder on the sphere, with wide-ranging applications.

\begin{figure*}
    \centering
    \includegraphics[width=0.96\textwidth]{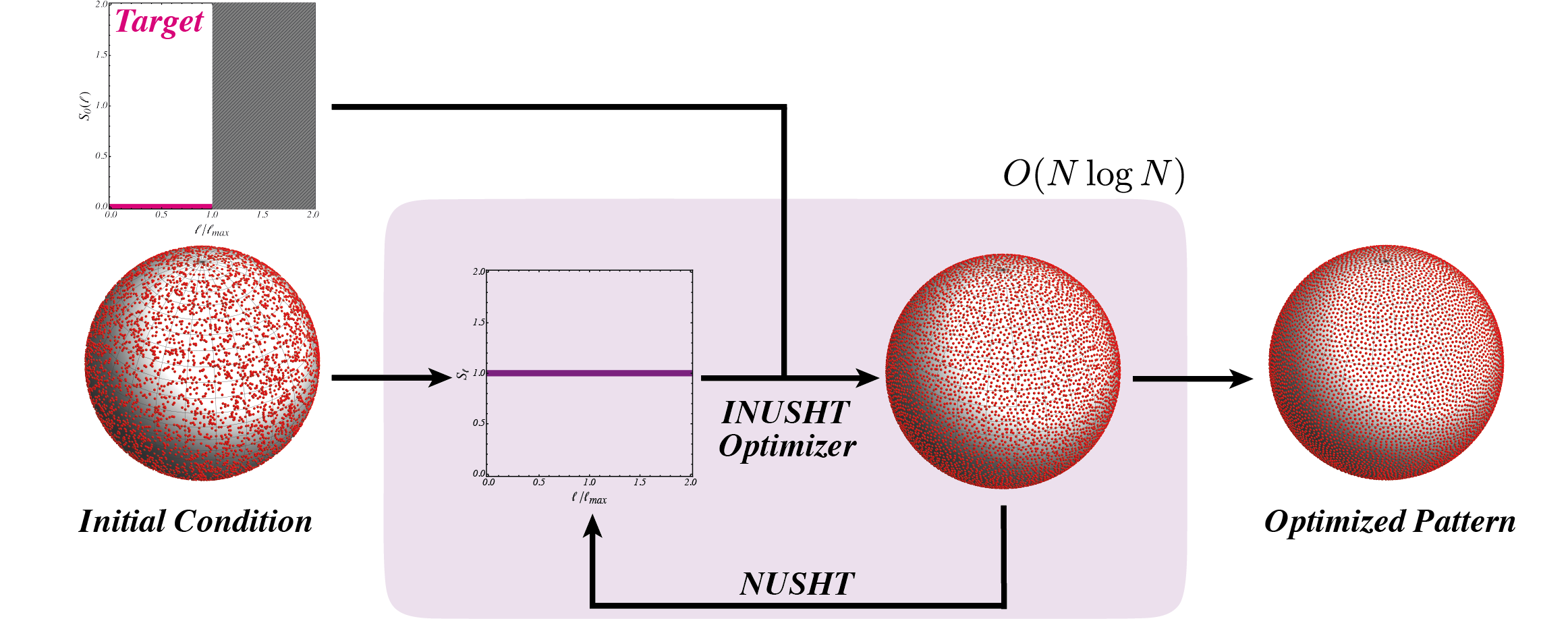}
    \caption{\textbf{Sketch of the algorithm.}
    We initialize the point pattern from a Poisson point pattern on the sphere and set a target over a finite amount of $\ell$ modes ($\ell \leq \ell_{\max}$).
    The power spectrum $S_\ell$ is computed using a NUSHT, then the gradients of the loss using the inverse transform (INUSHT), which is used to update positions within a gradient-based optimizer (here, L-BFGS).
    One such step takes $O(N \log N)$ operations.
    Steps are repeated until a convergence criterion is met, yielding the optimized pattern.}
    \label{fig:Algo}
\end{figure*}

\section{Setting}

We consider point patterns comprising $N$ points with positions $\bm{r}_1, \ldots, \bm{r}_N$ on the unit $2$-sphere $\mathcal{S} \subset \mathbb{R}^3$.
We represent this point pattern via a density field
\begin{align}
    \rho(\bm{r}) = \sum\limits_{n=1}^N c_n \delta\left(\bm{r} - \bm{r}_n\right),
\end{align}
where $c_n \in \mathbb{R}$ is the relative weight of point $n$ (with the convention $\sum_n c_n = N$) and the position of point $n$ can be parametrized as $\bm{r} = (\sin\theta\cos\phi, \sin\theta\sin\phi, \cos\theta)$ where $\theta\in[0;\pi]$ is the colatitudinal coordinate and $\phi\in[-\pi;\pi)$ the azimuthal one.

On the sphere, the relevant transform to consider for spectral optimization is not the Fourier Transform (FT) but the Spherical Harmonics Transform (SHT).
Indeed, spherical harmonics form a complete orthonormal basis of square-integrable ($L^2$) functions on the sphere, so that $\forall f\in L^2(S_2\mapsto \mathbb{C})$, one may write
\begin{align}
    f(\theta, \phi) = \sum\limits_{\ell = 0}^\infty \sum\limits_{m= -\ell}^\ell \wideparen{f}_{\ell}^m Y_{\ell}^m(\theta, \phi)
\end{align}
with 
\begin{align}
    \wideparen{f}_{\ell}^m \equiv \int\limits_{0}^{2\pi}d\phi \int\limits_{0}^\pi d\theta\sin\theta f(\theta,\phi) {Y_{\ell}^m}^\star(\theta,\phi),
\end{align}
where the star indicates a complex conjugate, and $Y_{\ell}^m$ is a spherical harmonic with indices $\ell \in \mathbb{N}$ and $-\ell \leq m\in\mathbb{Z} \leq \ell$.

With these conventions (see also App.~\ref{app:conventions}), the SHT of the density field reads
\begin{align}
    \wideparen{\rho}_\ell^m
    &= \sum\limits_{n=1}^N c_n {Y_{\ell}^m}^\star(\theta_n, \phi_n),\label{eq:rho_SHT}
\end{align}
which allows to define the power spectrum, or structure factor,
\begin{align}
    S_{\ell} \equiv \frac{4\pi}{N} \frac{1}{2\ell + 1} \sum\limits_{m=-\ell}^\ell\left |\wideparen{\rho}_{\ell}^m \right|^2.
\end{align}
This power spectrum encodes the pair correlations of the point pattern, analogously to its Fourier-space equivalent for Euclidean space~\cite{Pilleboue2015}.
In particular, for unit weights, $0\leq S_\ell \leq N$ and a set of uncorrelated random points drawn uniformly on the sphere (a Poisson point pattern) verifies $\langle S_\ell\rangle = 1$, where brackets indicate an average over realizations.
As a result, values $S_\ell <1$ can be interpreted as suppressed (sub-Poissonian) fluctuations and values $S_\ell>1$ as enhanced (super-Poissonian) fluctuations.
The goal in the following is to generate point patterns that realize a target $S_0(\ell)$ for some set of indices $\ell \in \Lambda \subset \mathbb{N}$.

\section{Algorithm}
To achieve such generation, inspired by the collective variable approach~\cite{Uche2004,Uche2006}, we introduce a loss function for the power spectrum,
\begin{align}
    L[S,S_0] = \sum\limits_{\ell\in \Lambda} W_{\ell} \left[ S_\ell - S_{0}(\ell) \right]^2,\label{eq:Loss}
\end{align}
where $W_\ell$ is a non-negative weight.
The key to the algorithm we introduce in this paper is that, much like the FReSCo loss~\cite{Shih2023,FReSCo} in flat space, both the loss in Eq.~\eqref{eq:Loss} and its gradients with respects to either weights or angular coordinates can be expressed as (direct or inverse) SHTs, see the sketch in Fig.~\ref{fig:Algo}.
Computing the loss itself amounts to computing the power spectrum, which we already showed can be computed using SHTs.
The full computation of the gradients is shown in App.~\ref{app:Gradients}.
We here only reproduce the final result for the derivative with respect to weights,
\begin{align}
    \frac{\partial L}{\partial c_p} &= \text{Re}\left[\sum\limits_{\ell \in \Lambda} \sum\limits_{m= - \ell}^\ell \wideparen{C}_{\ell}^m Y_{\ell}^m{} (\theta_p, \phi_p)\right] \label{eq:Loss_gradient_c_main} \\
    \wideparen{C}_\ell^m &\equiv  \frac{16\pi}{(2\ell + 1)N}   W_\ell (S_\ell - S_0(\ell)) \wideparen{\rho}_{\ell}^m{} \label{eq:SHT_cp_grad_main}
\end{align}
with respect to azimuthal angles,
\begin{align}
    \frac{\partial L}{\partial \phi_p} &= \frac{16\pi c_p}{N}\sum\limits_{\ell \in \Lambda} \sum\limits_{m=-\ell}^\ell \wideparen{\mathcal{D}}_\ell^m Y_{\ell}^m (\theta_p, \phi_p) \\
    \wideparen{\mathcal{D}}_\ell^m &\equiv i m\frac{W_\ell (S_\ell - S_0(\ell))}{2\ell+1} \wideparen{\rho}_{\ell}^m{}.
\end{align}
and with respect to the co-latitude,
\begin{align}
    \frac{\partial L}{\partial \theta_p} &= \frac{8\pi c_p}{N} \cos\phi_p\sum\limits_{\ell \in \Lambda} \sum\limits_{m=-\ell}^\ell \wideparen{\mathcal{F}}_{\ell}^m Y_{\ell}^m (\theta_p, \phi_p) \nonumber \\
    &\quad\quad+ \frac{8\pi c_p}{N} \sin\phi_p\sum\limits_{\ell \in \Lambda} \sum\limits_{m=-\ell}^\ell \wideparen{\mathcal{G}}_{\ell}^m Y_{\ell}^m (\theta_p, \phi_p),
\end{align}
with
\begin{align}
    \wideparen{\mathcal{F}}_{\ell}^m &\equiv \frac{W_\ell (S_\ell - S_0(\ell))}{2\ell+1} \left[  A_{\ell,m}\wideparen{\rho}_{\ell}^{m-1}{} - B_{\ell,m}\wideparen{\rho}_{\ell}^{m+1}{}\right], \\
    \wideparen{\mathcal{G}}_{\ell}^m &\equiv -i \frac{W_\ell (S_\ell - S_0(\ell))}{2\ell+1} \left[  A_{\ell,m}\wideparen{\rho}_{\ell}^{m-1}{} + B_{\ell,m}\wideparen{\rho}_{\ell}^{m+1}{} \right]
\end{align}
where $A_{\ell,m} = \sqrt{(\ell -m+1)(\ell+m)}$ and $B_{\ell,m} = \sqrt{(\ell +m+1)(\ell-m)}$.

In summary, optimizing positions or weights only requires 3 or 1 inverse SHTs per gradient evaluation, respectively, plus 1 direct SHT to evaluate $\wideparen{\rho}$ which is used throughout the loss and gradient evaluations.

The precise choice of SHT algorithm is then crucial to the speed of the overall optimization.
In flat space, past work~\cite{Shi2023,Casiulis2024a} highlighted that using multithreaded implementations of non-uniform fast Fourier transforms (NUFFTs) like finufft~\cite{Barnett2019,Barnett2020fixed} provided both high speed and bounded error on the Fourier transforms of point patterns, paving the way for fast, flexible, and reliable optimization.
By analogy, on the sphere, one needs high-performance non-uniform spherical harmonics transforms (NUSHTs).
We opted for the implementation provided in the Distinctly Useful Code Collection (ducc)~\cite{ducc}, an augmented spin-off of the libsharp library~\cite{Reinecke2013} that implements a number of additional speedups and memory optimizations, notably Ishioka's recurrence formula for efficient computation~\cite{Ishioka2018}.

Having chosen an SHT method, we only need to specify an optimization method to minimize the loss.
A subtlety compared to flat space is that fully rigorous Riemannian optimization would need to follow geodesics, here great circles of the sphere.
While such Riemannian optimization is possible, both on the sphere~\cite{Gräf2011} and on more general manifolds (see \textit{e.g.} Refs.~\cite{Seibert2013,Huang2018}), we use a simpler approach. 
Limiting ourselves to small steps (much smaller than $\pi$), we update positions following the gradient in Cartesian coordinates in the tangent space then project the resulting point back onto the unit sphere by normalizing the position vector, a common approach in spherical geometry (see, \textit{e.g.}, Refs.~\cite{Vest2014,Lei2023}).
Throughout this work, following Ref.~\cite{Shih2023}, the tangent-space step is proposed using an implementation of the Limited-Memory Broyden–Fletcher–Goldfarb–Shanno (L-BFGS) algorithm~\cite{Liu1989}, a quasi-Newton method that only uses gradient information but estimates the local Hessian on the fly with a finite memory of past gradient values.
Optimization is terminated either when the gradient falls below a set value, or when a maximum number of steps has been performed.
We provide a more complete description of the optimization method in App.~\ref{app:Optimization}, and share~\fashion~as a publically available library~\cite{fashion-pop}.

\section{Results}

Armed with the \fashion~algorithm, we now apply it to a few targets, each motivated by past works either on spheres or in Euclidean space.

\subsection{Stealthy Hyperuniformity}

A first example of target property is hyperuniformity~\cite{Torquato2018}, \textit{i.e.} the anomalous suppression of long-range density fluctuations.
As recently noted in the physics community~\cite{Meyra2019,LosdorfeerBozic2019,Lei2023}, and analogously to flat-space hyperuniformity, the spherical version of hyperuniformity is associated to the property $S_\ell \underset{\ell\to0}{\to}0$.
A special case of hyperuniformity is stealthy hyperuniformity, characterized by the property $S_\ell \approx 0$ for $\ell \leq \ell_{\max}$~\footnote{As argued in Ref.~\cite{Shih2023} in the case of Euclidean space, in practice, a reasonable criterion is $S_\ell  \lesssim 1/N$.}.
This kind of system has been the focus of many studies in flat space~\cite{Florescu2009,Man2013,Torquato2018,Cheron2022,Monsarrat2022,Shih2023,Siedentop2024}, but spherical stealthy point patterns have been relatively little studied by physicists.

Interestingly, such point patterns on the sphere have attracted considerable attention in mathematics, under the name of \textit{spherical $t$-design}~\cite{Gräf2011,Womersley2018}.
A spherical $t$-design is a point pattern on the sphere that, when it is used to evaluate the integral of a function as a discrete sum, yields the exact value for any spherical polynomial with degree lower than or equal to $t$.
The link between this definition and stealthy hyperuniformity is made clear by a known result in sampling~\cite{Pilleboue2015}: consider a function $f$ on the sphere, and stationary random point patterns with an average power spectrum $S_\ell$.
The mean-square error of the Monte Carlo estimator of the integral of $f$ obtained using these point patterns can be shown~\cite{Pilleboue2015} to verify
\begin{align}
    MSE \propto \sum\limits_{\ell = 1}^\infty S_\ell \mathcal{P}_\ell[f]
\end{align}
where $\mathcal{P}_\ell[f]$ is the $\ell$-th mode of the power spectrum of $f$.
Stating that $f$ is a spherical polynomial of degree $t$ means that its power spectrum is band-limited with $\mathcal{P}_\ell[f] = 0$ if $\ell > t$.
As a result, it is clear that a spherical $t$-design is obtained if a point pattern verifies $S_{\ell} = 0$ for $\ell \leq t$, so that one may get $MSE = 0$.

This also sheds light on the use of hyperuniform point patterns: they minimize error on Monte Carlo estimators for integrals, and thus constitute an optimal sampling scheme for band-limited signals on the sphere.
More generally, $t$-designs are relevant to optimal sampling for signal reconstruction~\cite{Potts2017,Chen2018a,Zhou2018a_fixed,Lin2024,Xiao2025}, in particular for physically based rendering~\cite{Pharr2018}, experimental acoustic signal measurements~\cite{Rafaely2005,Pinardi2021,Tanaka2025}, their analogues for electromagnetic waves~\cite{Marantis2009} or geophysical elastic waves~\cite{Chao2014}, numerical stencil generation for PDE integration on the sphere~\cite{Flyer2012} relevant to geophysical simulations, in particular atmospheric models~\cite{Ishioka2018}.

That is why the fast generation of spherical hyperuniform patterns has been the focus of some work~\cite{Balzer2009,Gräf2011,Womersley2018}.
However, past implementations suffer from high computational costs, with some works explicitly quoting computing times on the order of days to generate systems of $10^4$ points~\cite{Gräf2011}.
It is thus a perfect example to test and benchmark~\fashion.

\begin{figure}
    \centering
    \includegraphics[width=0.96\columnwidth]{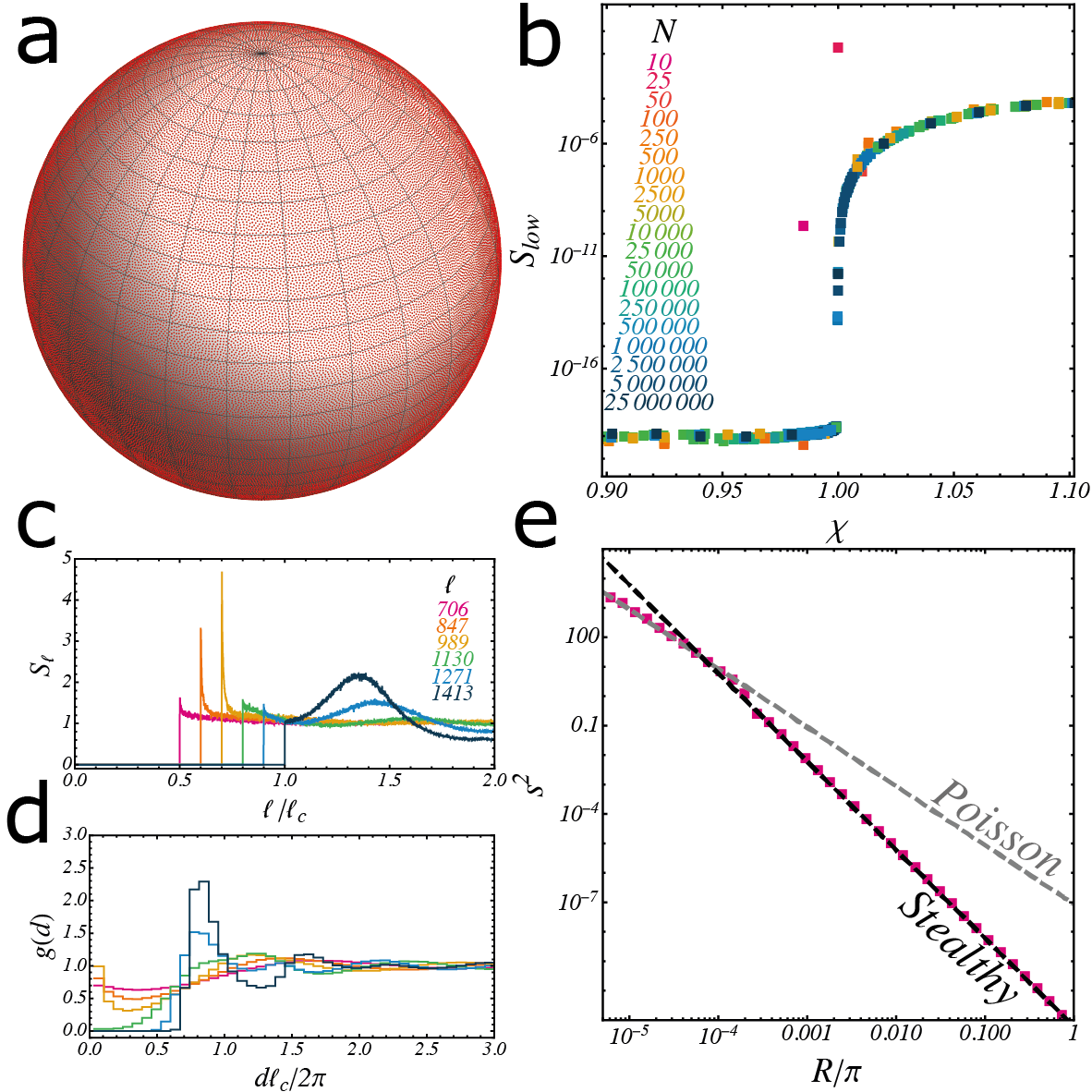}
    \caption{\textbf{Stealthy hyperuniformity on the sphere.}
    $(a)$ Example stealthy hyperuniform pattern on the sphere, obtained for $N = 10^5$ points and $\ell_{max} = 446$, corresponding to $\chi \approx 0.999$.
    $(b)$ Arithmetic average $S_{low}$ of the final $S_{\ell}$ in the constrained region against $\chi$, across system sizes.
    $(c)$ Final power spectra for $N = 10^6$ points and a few $\ell_{max}$ values given in inset.
    The $x$ axis is rescaled by $\ell_c$, the value of $\ell_{\max}$ that yields $\chi = 1$.
    $(d)$ Corresponding (spherical) radial distribution functions at a distance $d$.
    The $x$ axis is rescaled to have the first feature around $x = 1$.
    $(e)$ Reduced variance $s^2 = \text{var}(n_R) / \mathbb{E}[n_R]^2$ against the radius $R$ of the random spherical caps used in the calculation (here rescaled by the maximal value of $R$, namely $\pi$).
    Dashed lines indicate the Poisson scaling $s^2 \propto 1/R^2$ and the stealthy scaling $s^2 \propto 1/R^3$.
    The data was obtained by sampling uniformly the centers of $10^4$ spherical caps with radius $R$ in a system of $N = 5\times10^6$ and $\ell_{\max} = 3159$ ($\chi \approx 0.99$).}
    \label{fig:SHU}
\end{figure}

We first show results regarding the nature of the obtained point patterns in Fig.~\ref{fig:SHU}.
In Fig.~\ref{fig:SHU}$(a)$, we show an example snapshots of generated SHU patterns such that $S_\ell = 0$ for $\ell \leq \ell_{\max}$ with $N = 10^5$ and $\ell_{\max} = 446$.
Note that the point pattern exhibits the characteristic texture of stealthy hyperuniform point patterns~\cite{Uche2004,Torquato2018,Shih2023}.

To better characterize the degree of hyperuniformity of structures, following classical arguments from the field of collective variables~\cite{Uche2004,Uche2006}, we resort to a Maxwell-type~\cite{Maxwell1864} degree-of-freedom counting and introduce
\begin{align}
\chi \equiv \frac{\#(\text{constraints})}{\#(\text{available dof})} = \frac{(1+\ell_\text{max})^2-1}{2(N-1)}.
\end{align}
In this expression, the number of constraints is evaluated by counting all $(\ell,m)$ pairs that are implicitly constrained, noticing that the $\ell = 0$ mode is trivially always set by $N$, and the number of degrees of freedom is obtained by noticing that the SHT representation is translation-invariant on the sphere so that the center-of-mass position is not an available degree of freedom.
Naïvely, from past work on SHU systems~\cite{Uche2004,Uche2006,Torquato2018,Shih2023}, one would expect optimization to a disordered SHU structure to be possible as long as $\ell_{\max} \leq \ell_c$ with a critical value set by $\chi(\ell_{c}) = 1$~\footnote{We here voluntarily use a definition for $\chi$ such that saturation happens at $1$, although some past works in flat space have used a definition that differs by a factor of 2~\cite{Uche2004,Uche2006,Torquato2018,Shih2023}}.
In Fig.~\ref{fig:SHU}$(b)$, we report the value of the arithmetic mean of $S_\ell$ in the constrained region, $S_{low} \equiv \ell_{\max}^{-1} \sum_{\ell =1}^{\ell_{\max}} S_\ell$, as a function of $\chi$ for values of $N$ going from $10$ to $2500000$.
We find that the Maxwell-type argument is in remarkably good agreement with a realizability threshold, analogous to a SAT-UNSAT transition in constraint satisfaction~\cite{Krzakala2007}, at $\chi = 1$.
This constitutes conclusive evidence regarding the ongoing debate~\cite{Womersley2018} on the maximal value of $t$ in a $t$-design as a function of $N$, which to the best of our knowledge is not settled by an analytical argument.

To illustrate the buildup of order with $\chi$ for $\chi \leq 1$, we show a few power spectra at a constant $N = 10^6$ but growing $\chi$ in Fig.~\ref{fig:SHU}$(c)$, and the corresponding radial distribution functions $g$ as a function of the great-circle distance $d$ in Fig.~\ref{fig:SHU}$(d)$ (see App.~\ref{app:PairCorr} for the definition of the rdf on the sphere).
We show that, like in flat space, all these configurations are disordered, in the sense that both $S_\ell$ and $g(d)$ relax to the Poisson value, $1$, as $\ell$ and $d$ grow.
Like in flat space, $S_\ell$ displays a wide ripple outside of the constrained region that grows with $\chi$~\cite{Shih2023,Casiulis2024a}.
Even values of $\chi$ near the realizability transition yield remarkably little structure in $g(d)$, with only 2 notable peaks in the data we show.
Also note that the $g(d)$ develops a sharper and sharper exclusion radius as as $\ell$ grows, such that $\chi \to 1^-$ corresponds to strict hardcore exclusion for a finite repulsive diameter.

Finally, we assess the quality of the stealthiness through ``discrepancy'' measurements, \textit{i.e.} point count fluctuations in balls with random, uniformly drawn centers~\cite{Torquato2018}.
Like in flat space~\cite{Shih2023}, one expects the variance $\text{var} \,n_R$ of the number of points in a window to grow abnormally slowly with the size $R$ of the window, $\text{var}\, n_R \sim R$ instead of the Poissonian scaling $\text{var}\, n_R \sim R^2$ (this is a special case of the more general result on the MSE of a point pattern in a Monte Carlo integration~\cite{Pilleboue2015}, here applied to a spherical cap).
Following standard conventions from the SHU literature~\cite{Torquato2018,Shih2023}, we introduce $s^2 = \text{var} \, n_R / \mathbb{E}[n_R]^2$ the reduced variance and check its scaling against $R$ in Fig.~\ref{fig:SHU}$(d)$.
We show that our point patterns display a crossover between a Poissonian scaling $s^2 \sim R^{-2}$ at small $R$ and a $s^2 \sim R^{-3}$ at large $R$, across four decades of $R$, thus indicating clear stealthy hyperuniformity in a real-space measurement. 

\begin{figure}
    \centering
    \includegraphics[width=0.96\columnwidth]{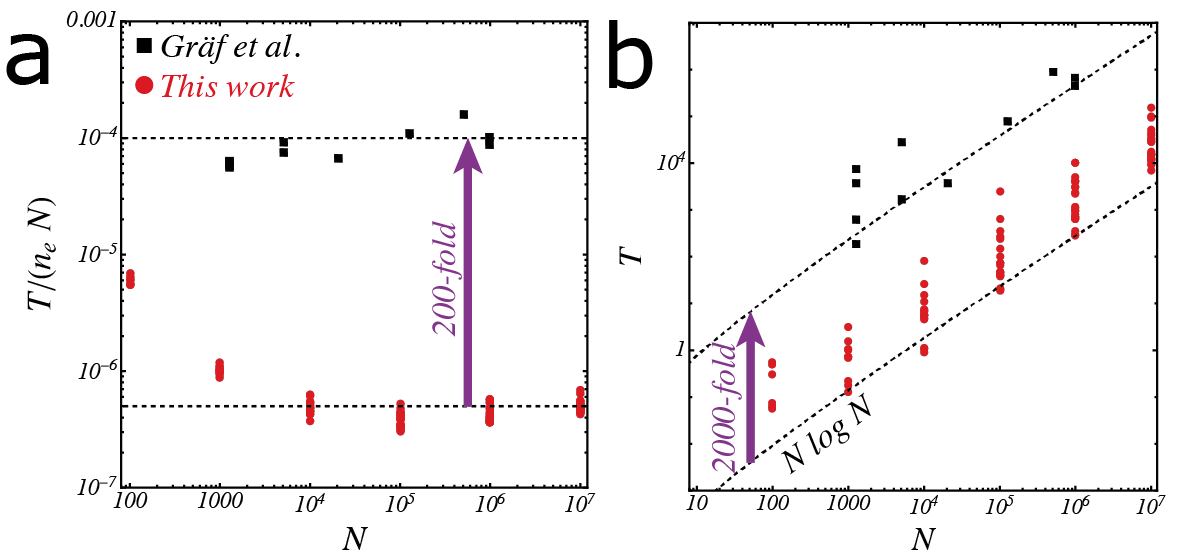}
    \caption{\textbf{Benchmark.}
    $(a)$ Total optimization time $T$, in seconds, divided by the number of gradient evaluations $n_e$ and by the number of particles $N$.
    Our data (red symbols) was obtained across values of $\chi$ from $0.5$ to $1$ at each $N$.
    Black symbols represent times reported in Ref.~\onlinecite{Gräf2011}.
    We highlight the asymptote of~\fashion~data by a dashed line, and a plausible line going through Gräf's and Pott's data~\cite{Gräf2011}.
    $(b)$ Raw time, in seconds, for the same data.
    We highlight with dashed lines plausible $N \log N$ scaling that capture the lower-end of both datasets.
    }
    \label{fig:SHU_benchmark}
\end{figure}

Having shown that~\fashion~produces convincingly SHU structures across scales, we now turn our attention to its algorithmic performance in the example of SHU, which is the only one with documented time scalings in the literature on spherical point pattern optimization~\cite{Gräf2011}.
To assess the performance of the algorithm, we restrict ourselves to the $\chi < 1$ region, since it is the only one in which convergence to an actual minimum of the loss is guaranteed.
We measure the wall time $T$ taken by the whole optimization, as well as the number of evaluations $n_{e}$ required to converge, across $N$ and $\chi$.
We report the results in Fig.~\ref{fig:SHU_benchmark}.

First, in Fig.~\ref{fig:SHU_benchmark}$(a)$, we show the scaling of the time per evaluation and per particle, $T / (N n_e)$, across values of $N$ and $\chi \leq 1$.
We report both our own values (red symbols) and values reported in Ref.~\onlinecite{Gräf2011}.
We note that the time per evaluation and per particle of our algorithm converges to a constant at large $N$, while past values seem to display a very slow growth.
Furthermore, we report that the constant asymptote of our algorithm saturates at a value hundreds of times lower than in the past implementation.
The difference gets even starker when plotting the total optimization time in seconds, shown in Fig.~\ref{fig:SHU_benchmark}$(b)$.
Showing all of our data and all of the disclosed times from Ref.~\onlinecite{Gräf2011}, we report that the scaling of the lower bound of each group of points is compatible with an $N \log N$ scaling, but with a speed-up that is on the order of thousands.
Note in particular that we use values of $\chi$ spanning from $0.5$ to $1.0$ at all tested values of $N$, while Gräf et al. only considered $\chi \approx 0.5$ at large $N$.
Furthermore, note that we only ever start from Poisson point patterns while they also considered optimizations started from Fibonacci spirals~\cite{Gräf2011}, and that our final average structure factor, $S_{low}$, reported in Fig.~\ref{fig:SHU}, is lower than theirs.
Our algorithm would thus likely be even faster in absolute time if the comparison were fully fair~\footnote{In particular, our computation times are better even taking into account the difference in hardware. Our computations were performed on 24-core Intel Cascade Lake Platinum 8268 chips (2.9 GHz, up to 3.9 GHz in turbo mode; 24-core), while Gräf et al.'s were performed on a 4-core Intel Core i7 CPU 920 processor~\cite{Gräf2011} (2.66 GHz). Boldly assuming that the computation fully exploits the differences between frequencies and number of cores, the expected ratio of times would be less than 10, not in the hundreds or thousands -- even taking into account a very optimistic 50\% speed-up in per-core computation speed.}.

Having shown that~\fashion~performs better than past algorithms that implement spectral optimization on the sphere, we now demonstrate that it also comes with a large flexibility in the choice of targets, much like FReSCo~\cite{Shih2023} -- a notable difference from algorithms that are highly specialized to the production of stealthy hyperuniform structures~\cite{Balzer2009,Gräf2011}.

\subsection{Power-law Hyperuniformity}

\begin{figure}
    \centering
    \includegraphics[width=0.98\columnwidth]{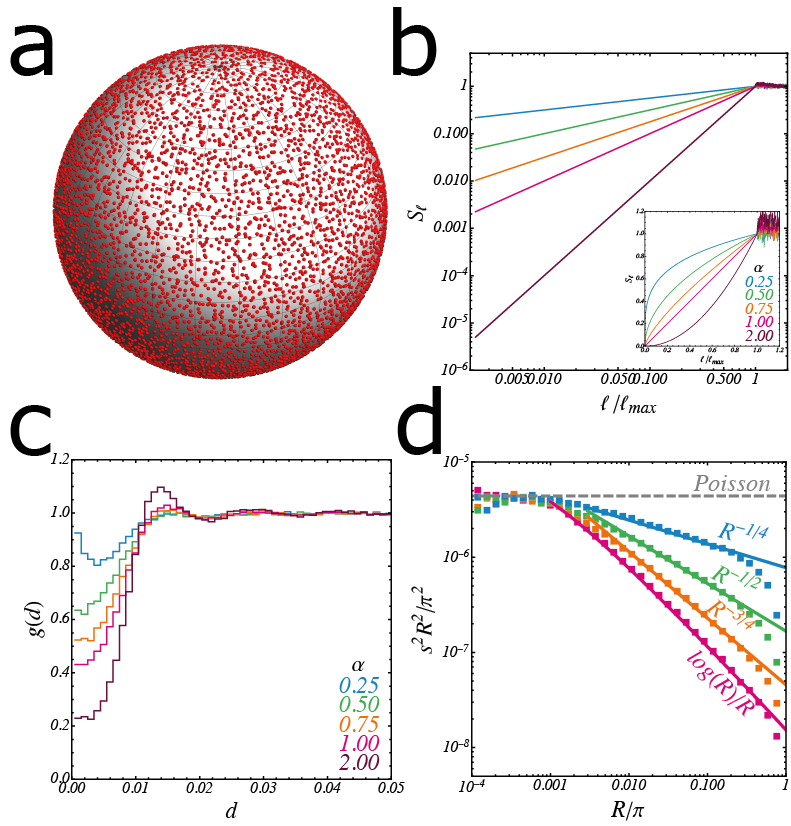}
    \caption{\textbf{Power-law hyperuniformity.}
    $(a)$ Example optimized structure for $N=10^4$ points with a power law $S_\ell \propto \ell$ up to $\ell_{\max} = 140$ ($\chi \approx 1$).
    $(b)$ Final power spectra $S_\ell$ in log-log scale for $N = 10^5$ for constraints $S_\ell \propto \ell^\alpha $ up to $\ell_{\max} = 446$ ($\chi \approx 1$) for $\alpha$ values $0.25, 0.5, 0.75, 1,2$.
    Inset: linear-scale plots of the same curves.
    $(c)$ Corresponding radial distribution functions.
    $(d)$ Corresponding discrepancy measurements, here rescaled as $s^2 R^2/\pi^2$ to highlight $R^{-\alpha}$ decays (continuous lines), and the $\log(R)/R$ decay for $\alpha = 1$.
    The $\alpha = 2$ case is here omitted to make the $\alpha = 1$ decay more visible.}
    \label{fig:PLHU}
\end{figure}

Another family of structures of interest are power-law hyperuniform ones, where $S_\ell \propto \ell^\alpha$ as $\ell \to 0$.
In flat space, analogous structures have been the focus of many works either because they are indicative of criticality in absorbing-state phase transitions~\cite{Hexner2015,Hexner2017,Dandekar2020,Wilken2021,Wiese2024,Mukherjee2024,Jiao2011a,Dam2025,Wang2025a,Pine2005,Corte2008,Wilken2020,Weijs2015} or of structured noise in non-equilibrium steady states~\cite{Lei2024}, notably in active matter systems~\cite{Lei2019,Lei2019a,Huang2021,Chen2024,Kuroda2025,Kuroda2025a,Zhang2022,DeLaCotte2025,Wang2025}, diffusive processes~\cite{Jack2015,Ikeda2023,Ikeda2025}, driven rotors~\cite{Oppenheimer2022}, and shaken granulars~\cite{Maire2025a}.

On the sphere, only recently did physicists take an interest in this kind of hyperuniformity~\cite{Lei2023}, and much is yet to learn about the realizability and occurrences of spherical power-law hyperuniformity.
We thus showcase that~\fashion~may be used to produce such structures without introducing~\textit{ad hoc} dynamics.
Results are shown in Fig.~\ref{fig:PLHU}.

In Fig.~\ref{fig:PLHU}$(a)$, we illustrate a power-law hyperuniform pattern on the sphere, using the example of an $N=10^4$ system with a linear constraint, $S_0(\ell) = \ell/\ell_{\max}$, such that $\chi \approx 1$.
Note in particular that the pattern is notably less homogeneous than the SHU example of Fig.~\ref{fig:SHU}.
In Fig.~\ref{fig:PLHU}$(b)$, we plot the final $S_\ell$ after optimization for a few exponents $\alpha$ of power-law targets $S_0(\ell) = (\ell / \ell_{\max})^\alpha$, in log-log (main panel) and linear (inset) scales.
To characterize their order in real space, we also show their radial distribution functions in Fig.~\ref{fig:PLHU}$(c)$, highlighting that a higher exponent leads to a higher degree of exclusion at short range.
Finally, we check in Fig.~\ref{fig:PLHU}$(d)$ that these exponents are associated to the expected decays of the discrepancy, $R^2 s^2 \sim R^{-\alpha}$ for $\alpha < 1$ (``class III'' hyperuniformity~\cite{Torquato2018}) and $R^2 s^2 \sim \log (R) /R$ for $\alpha = 1$ (``class II'' hyperuniformity~\cite{Torquato2018}).
Higher powers, even the SHU case which corresponds to $\alpha \to \infty$, display the maximally suppressed fluctuations $R^2 s^2 \sim 1/R$ for $\alpha = 1$ (``class I'' hyperuniformity~\cite{Torquato2018}), see Fig.~\ref{fig:SHU}$(e)$.
Altogether, these results show that~\fashion~achieves flexible and precise optimization for power-law targets, not just optimal computational speed.
Furthermore, we have shown that the classification of hyperuniform point patterns on the plane was directly extendable to the sphere with the same limiting cases for discrepancy behavior.

\subsection{Gyromorphs}

Like FReSCo~\cite{Shih2023},~\fashion~is a promising strategy to design disordered metamaterials for transport properties, here patterned spheres -- which have recently been explored as a promising platform for structural color~\cite{Manne2024}.
As a minimal example of disordered metamaterials, we consider the example of gyromorphs~\cite{Casiulis2024a}, structures introduced in $2d$ flat space that are optimized to display a ring of high peaks in reciprocal space and a minimal separation in real space.
These structures were shown to display wide bandgaps both for photonic~\cite{Casiulis2024a} and electronic~\cite{Paz2026} transport.
We extend their definition to the sphere as point patterns such that $S_\ell$ is maximized for a particular $\ell$ value but is left unconstrained elsewhere.

We also adapt the algorithm in the same way as FReSCo between Refs.~\cite{Shih2023} and~\cite{Casiulis2024a} to avoid point overlaps, namely by cycling phases of optimization and ones of point removals and farthest-first batch insertions~\cite{Rosenkrantz1977}, see App.~\ref{app:Optimization}.

\begin{figure}
    \centering
    \includegraphics[width=0.98\columnwidth]{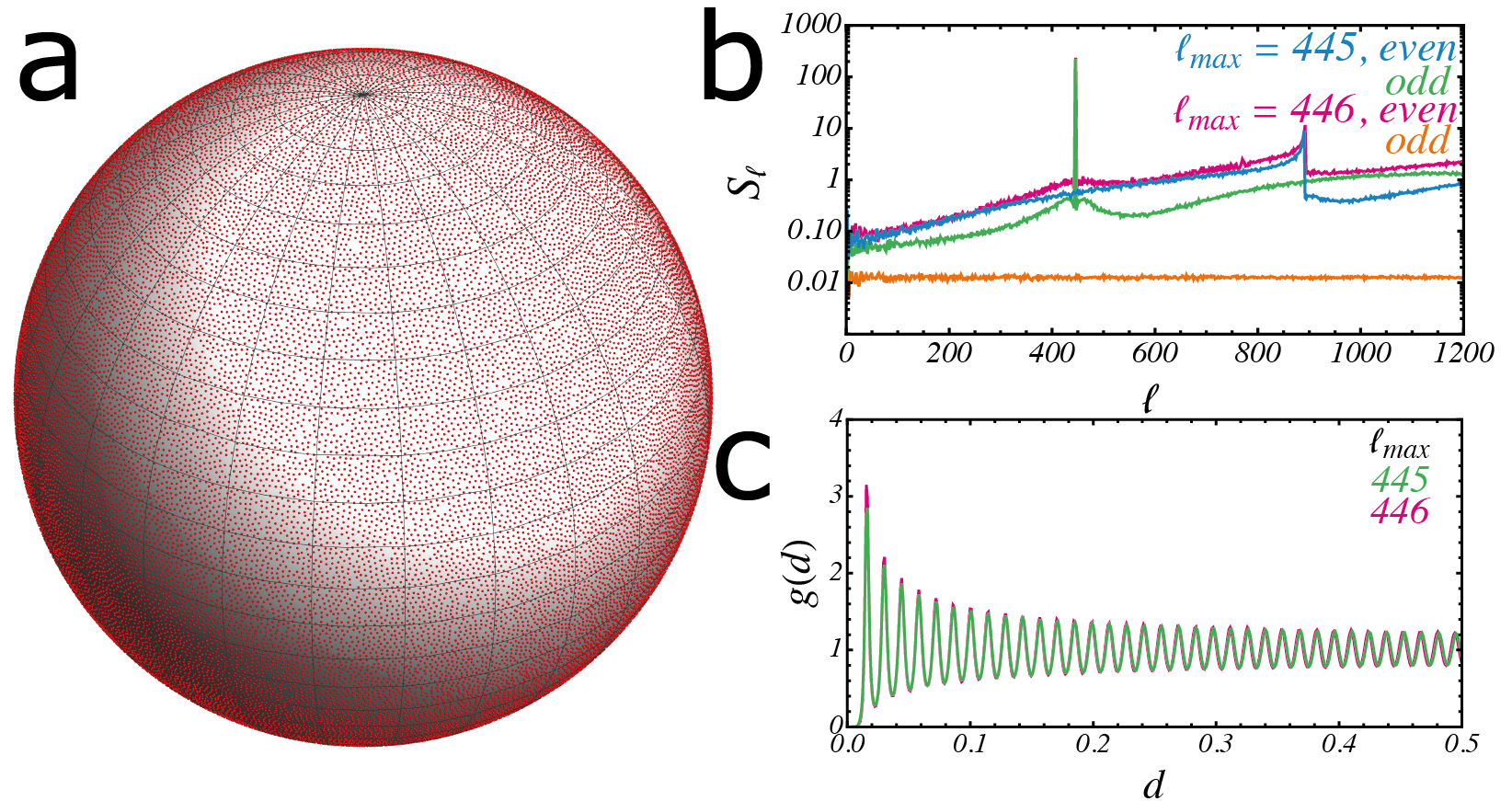}
    \caption{\textbf{Gyromorphs.}
    $(a)$ Example gyromorph with $N \approx 10^5$ and $\ell_{\max} = 446$.
    $(b)$ Log-scale power spectrum of optimized structures with $N \approx 10 ^5$ for $\ell_{\max} = 445$, split between the even (blue) and odd (green) $\ell$ values, and for $\ell_{\max} = 446$, split between the even (red) and odd (orange) $\ell$ values.
    $(c)$ Corresponding radial distribution functions for $\ell_{\max} = 445$ (green) and $\ell_{\max} = 446$ (red), against the distance $d$ in radians.}
    \label{fig:Gyros}
\end{figure}

Results are shown in Fig.~\ref{fig:Gyros}.
In Fig.~\ref{fig:Gyros}$(a)$, we show an example point pattern obtained for $N = 10^5$ and maximizing the height of the peak at $\ell_{max} = 446$.
Notice the characteristic~\cite{Casiulis2024a} circles of far neighbors around each point.
In Fig.~\ref{fig:Gyros}$(b)$, we show the corresponding power spectrum, splitting the plot between the even (red) and odd (orange) values of $\ell$.
The reason for this split is that for even values of $\ell_{\max}$, the strutures spontaneously attenuate all odd modes $S_{2\lambda +1}$ with $\lambda \in \mathbb{N}$(orange line), encoding the fact that the resulting point pattern is near-centrosymmetric.
The even modes, $S_{2\lambda}$, on the other hand, is reminiscent of the structure factor of gyromorphs in flat space~\cite{Casiulis2024a}.
There is a high peak at $\ell_{\max}$ as expected, which is much higher than any other feature, with a much weaker echo at $2 \ell_{\max}$.
The rest of the structure is liquid-like: it is not hyperuniform at $\ell \to 0$, and $S_{2\lambda }\to 1 $ as $\ell \to \infty$.
Because of this split, we also show the power spectrum of a structure with an odd peak location, $\ell_{\max} = 445$.
For this choice, there is still a split between odd (green) and even (blue) values of $\ell$, as they follow distinguishable curves, yet both have non-trivial structure.
Finally, in Fig.~\ref{fig:Gyros}, we show the corresponding radial distribution functions for both values of $\ell_{\max}$, and thus show that the pair correlations are very similar, indicating that the difference between odd and even values of $\ell_{\max}$ is subtle to catch at short range in real space.

Amplifying specific modes in the spectrum of spherical point patterns paves the way for a systematic exploration of regular coverings of the sphere, as crystals and lattices also display strong peaks.
Furthermore, in the context of sampling, maximizing the height of a specific mode creates a pattern that amplifies one particular mode of a function sampled at these points -- which may be of use to optimally measure signals corresponding to high-order harmonics, for instance in geophysics.
Finally, flat-space gyromorphs display maximal scattering intensity at half the wave-vector of the peak, which is why they display good photonic~\cite{Casiulis2024a} and topological electronic~\cite{Paz2026} bandgaps, with possible applications to the design of freeform waveguides~\cite{Man2013a}.
Spherical-space gyromorphs should display the same property for waves confined on the sphere, with exciting possible applications, \textit{e.g.} to whispering gallery modes~\cite{Yu2021} or structural color~\cite{Manne2024}.

\subsection{Centrosymmetric Stealthy Coverings}

\begin{figure}
    \centering
    \includegraphics[width=0.96\columnwidth]{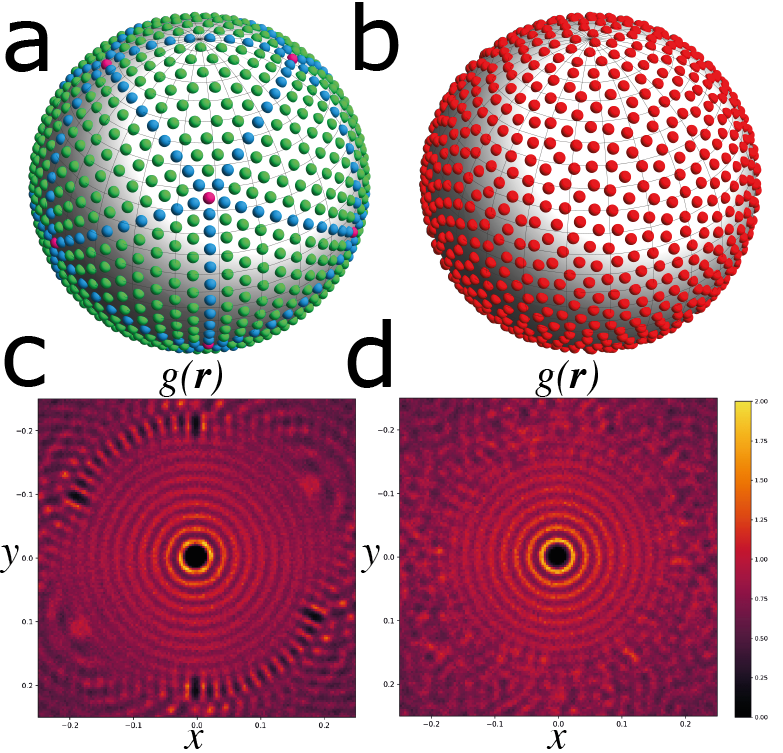}
    \caption{\textbf{Centrosymmetric SHU and $3d$ gyromorphs.}
    $(a)$ Peak pattern used in Ref.~\cite{Casiulis2024a} with $N = 1212$ peaks on a subdivided icosahedron, with vertices in red, edges in blue, faces in green.
    $(b)$ Example of a centrosymmetric SHU pattern with $N = 1212$ peaks.
    $(c)$ Slice of the pair correlation function $g(\bm{r})$ of a $3d$ gyromorph generated by the peaks of $(a)$, in the $(Oxy)$ plane.
    $(d)$ Likewise, but for a $3d$ gyromorph generated by the peaks of $(b)$.
    In both $(c)$ and $(d)$, lengths are in units of the total sidelength $L$ of the system.
    }
    \label{fig:centroSHU}
\end{figure}

We now highlight another axis of flexibility of~\fashion, the possibility to impose additional global symmetries in the point pattern.
Inspired by the emergent near-centrosymmetry of even-order spherical gyromorphs, Fig.~\ref{fig:Gyros}$(b)$, we consider the example of centrosymmetric point patterns, \textit{i.e.} ones such that the total number of points is even, $N = 2P$, and such that each point has an antipodal symmetric, so that $\forall n \leq P $ $\bm{r}_n = - \bm{r}_{n + P}$.
By optimization, one may achieve disordered point patterns with specific correlations, \textit{e.g.} hyperuniformity, and centrosymmetry.

The optimization scheme for centrosymmetric patterns is simple: the loss and its gradient are computed using the full point pattern, but only one half of the points are updated following it, while the other half is enslaved to their antipodes (see App.~\ref{app:CentroLoss}).
Amusingly, an example in which an optimized centrosymmetric covering of the sphere is desirable is the generation of $3d$ gyromorphs using FReSCo.
Indeed, in $3d$ gyromorphs, one selects $G/2$ points on a sphere with radius $K$ and maximize the value of the $3d$, Euclidean structure factor $S(\bm{k})$ at these peaks, which implicitly constrains their antipodes due to Friedel's law~\cite{Friedel1913,Schwarzenbach1996}.
Ideally, one also wants the peaks to cover the sphere uniformly and in a disordered way, so as to obtain an isotropic point pattern.
Thus, the ideal peak structure needs to be  disordered, uniform and centrosymmetric on the sphere.
In the seminal work on gyromorphs~\cite{Casiulis2024a}, because spherical optimization was unavailable, the authors opted for a deterministic subdivision of an icosahedron on the sphere, see Fig.~\ref{fig:centroSHU}$(a)$.
That subdivision had three issues.
First, only a sparse subset of integers was available for $G$, namely $G = 12 + 30 s + 10s(s-1)$ with $s$ the number of subdivision of each edge -- for instance, in Fig.~\ref{fig:centroSHU}$(a)$, $s =10$ leading to $G = 1212$.
Second, not all peaks are equivalent -- one may for instance notice in Fig.~\ref{fig:centroSHU}$(a)$ that the vertices of the original icosahedron have five neighbors while other points have six.
Finally, the overall system of peaks retained an overall icosahedral symmetry, affecting the degree of isotropy of the optimized gyromorphs.

Thus, as an illustration of the use of centrosymmetric optimized patterns, we propose to generate centrosymmetric stealthy hyperuniform patterns that can then be used as the locations of the peaks in a flat-space $3d$ gyromorph through the use of FReSCo.
Results are shown in Fig.~\ref{fig:centroSHU}.
In Fig.~\ref{fig:centroSHU}$(b)$, we show a centrosymmetric stealthy hyperuniform (cSHU) pattern with $N = 1212$ and $\ell_{\max} = 50$ ($\chi \approx 0.999$), to be compared with the pattern used in Ref.~\cite{Casiulis2024a}, shown in Fig.~\ref{fig:centroSHU}$(a)$.
It is clear that the cSHU pattern is much more disordered than the subdivided icosahedron, and that all points on it are equivalent up to random fluctuations due to the precise realization shown.
We then generate a $3d$ gyromorph from each of these peak structures, following the same method as in Ref.~\cite{Casiulis2024a} (which is identical to the one described in App.~\ref{app:Optimization} but using FReSCo in $3d$ space rather than~\fashion), and we show a $2d$ slice of their $3d$ pair correlation functions in Figs.~\ref{fig:centroSHU}$(c)$ and Fig.~\ref{fig:centroSHU}$(d)$.
This minimal example highlights that the structure of the peaks used for optimization yields a more isotropic structure in the subdivided icosahedron case, Fig.~\ref{fig:centroSHU}$(c)$: there are special axes in the plane, and even the ring of nearest neighbors is not fully isotropic.
By contrast, the gyromorph generated by a cSHU peak pattern looks completely isotropic, Fig.~\ref{fig:centroSHU}$(d)$.
Such a difference in the isotropy of local environments could have deep implications on the transport properties of the resulting structure, as it is known in photonics for instance that the uniformity of local environments is an important factor to achieve good photonic bandgaps~\cite{Sellers2017}.
We thus propose that cSHU generation via~\fashion~in the context of $3d$ gyromorphs is an interesting avenue for future works in materials design.

This centrosymmetric optimization has other uses, for instance in the context of ray-tracing techniques in rendering, in which one needs to spawn random rays on the hemisphere to compute Monte Carlo integrals of an illumination field on a camera plane~\cite{Pharr2018}, which is equivalent to generating centrosymmetric patterns on the full sphere.
Using spectrally optimized point patterns in that context could help produce optimal sampling for rendering, and thus alleviate computational costs for real-time rendering applications.

\section{Combining~\fashion~loss and pair repulsion}

\begin{figure}
    \centering
    \includegraphics[width=0.98\columnwidth]{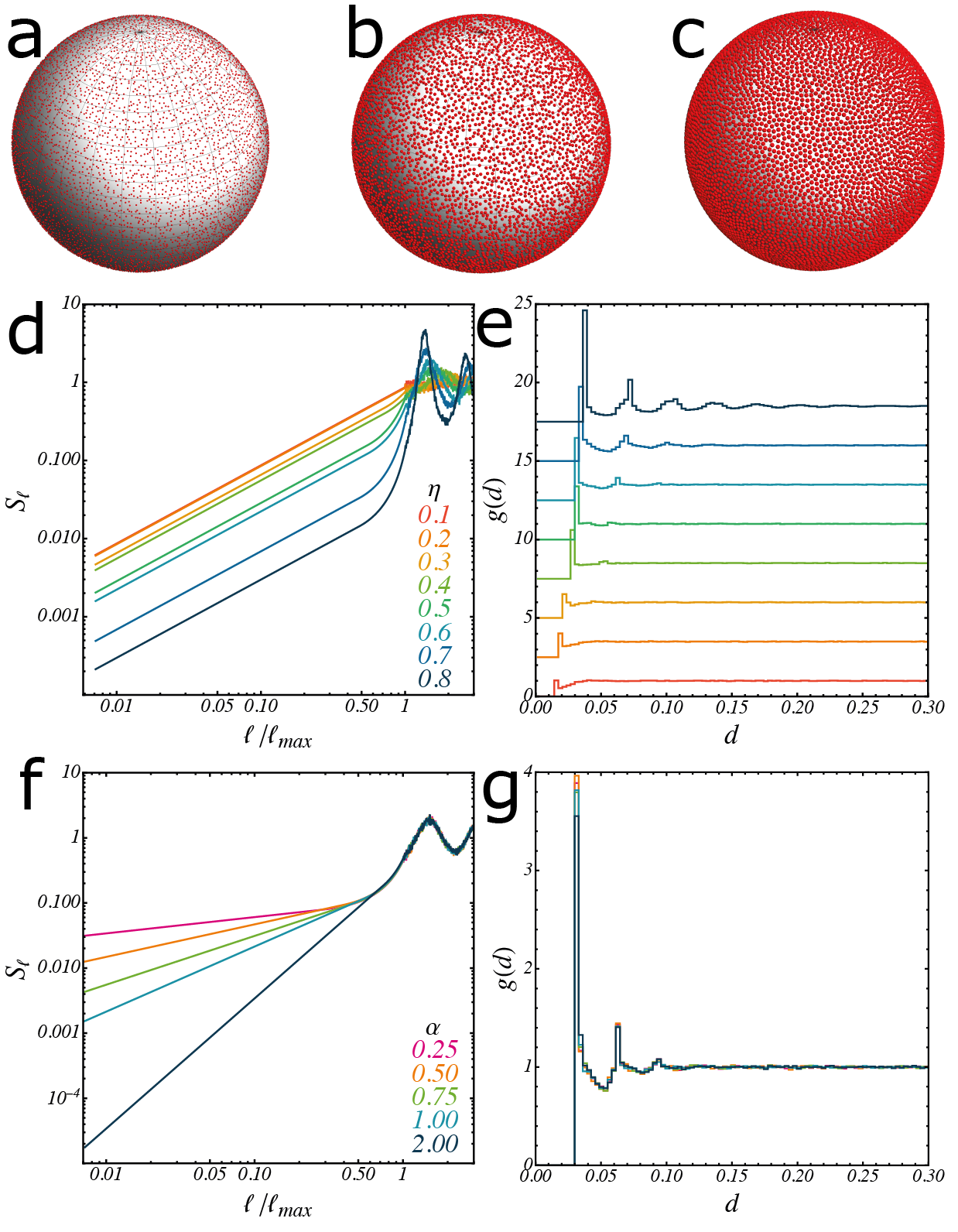}
    \caption{\textbf{\fashion~and repulsion.}
    $(a)-(c)$ Example optimized systems with $N = 10^4$, $S_\ell \propto \ell$ below $\ell_{\max} = 140$, ($\chi \approx 1$) and with Hertzian repulsion with packing fractions $(a)$ $\eta = 0.1$,$(b)$ $\eta = 0.4$ ,$(c)$ $\eta = 0.7$.
    Ball sizes represent the actual repulsive radii.
    $(d)$ Structure factors for systems with $N = 10^4$, $S_\ell \propto \ell$ below $\ell_{\max} = 140$ and across $\eta$, in log-log scales.
    $(e)$ Corresponding radial distribution functions (curves are here shifted for visibility)
    $(f)$ Structure factors for systems with $N = 10^4$, $\eta = 0.6$, and $S_\ell \propto \ell^\alpha$ below $\ell_{\max} = 140$, for a few exponents $\alpha$.
    $(g)$ Corresponding radial distribution functions.
    }
    \label{fig:Repulsive}
\end{figure}

Finally, much like FReSCo, the~\fashion~loss, Eq.~\eqref{eq:Loss}, can be summed with a standard pair repulsion, resulting in a total loss
\begin{align}
    L(\bm{r}_1, \ldots , \bm{r}_N) = \sum\limits_{\ell \in \Lambda} W_\ell \left[S_\ell - S_0(\ell) \right]^2 + \sum\limits_{n=1}^N \sum\limits_{p\neq n} U(\bm{r}_{m} - \bm{r}_n).
\end{align}
While the repulsion term naïvely looks like an $O(N^2)$ term, in the special case of finite-range repulsion with a range much smaller than $\pi$, using an appropriate division of the sphere into cells brings this cost down to $O(N \log N)$, like in flat space~\cite{Frenkel2001}.
As a result, one may in principle optimize a loss combining short-range real-space repulsion and long-range spectral constraints in quasilinear time.
We demonstrate this using the example of a power-law hyperuniform constraint combined with a finite-range potential (here, Hertzian repulsion $U(r) = \varepsilon(\sigma - r)^{5/2}$ if $r < \sigma$, with $\varepsilon$ a factor that sets the energy scale and $\sigma$ the repulsive range), using a cell list adapted to spherical geometry and described in detail in App.~\ref{app:CellList}.
Following Ref.~\cite{Shih2023}, in order to obtain smooth continuous power spectra at finite densities, the constraint is not just a power law at low $\ell$, but a function that smoothly interpolates between that power law at low $\ell$ and the power spectrum predicted by the spherical version of the Percus-Yevick equation of state~\cite{Lishchuk2006,Tarjus2012} at large $\ell$, see App.~\ref{app:PY}.

Results are shown in Fig.~\ref{fig:Repulsive}.
In Fig.~\ref{fig:Repulsive}$(a)-(c)$, we show a few examples of systems optimized with the same power law exponent but with different values of the filling fraction $\eta \equiv N (1 - \cos (\sigma/2))/2$.
These snapshots clearly show a lack of overlap between particles as expected, but also a globally isotropic pattern even at high $\eta$.
To go further, in Fig.~\ref{fig:Repulsive}$(d)$, we show the optimized power spectra for a linear power law ($\alpha =1$) but across filling fractions from $\eta = 0.1$ (dilute) to $\eta = 0.8$ (dense).
We show that the initial power law is convincingly imposed across density, and that it connects to a region at large $\ell$ with a growing degree of liquid-like structure~\cite{Hansen2006}.
This is confirmed in Fig.~\ref{fig:Repulsive}$(e)$ by the corresponding radial distribution functions (here shifted vertically for clarity), that show both a growing exclusion radius and an increasing degree of short-range structure as $\eta$ grows.
To complete the picture, in Fig.~\ref{fig:Repulsive}$(f)$, we show optimized power spectra, this time keeping $\eta = 0.6$ constant but varying $\alpha$ between $0.25$ and $2$.
We show that the power laws are convincingly imposed in spite of the (intermediate) filling fraction creating structure at large $\ell$.
Finally, in Fig.~\ref{fig:Repulsive}$(g)$, we show that the corresponding radial distribution function at short range is essentially unaffected by the choice of power law.
This confirms the intuition that the power law at low values of $\ell$ primarily affects long-range features of the point pattern, while repulsion primarily affects short-range ones.

In summary, we showed that we can convincingly impose non-trivial long-range properties like power-law hyperuniformity jointly with real-space constraints like pair repulsion on the sphere.
This approach may help understanding still little-studied examples of critical states of absorbing-state phase transitions on the sphere~\cite{Lei2023} in realistic cases of non-overlapping objects, and facilitate their manufacturing with real objects with finite sizes.
Furthermore, being able to design point patterns on the sphere that are simultaneously hyperuniform and well separated enough at short range has been argued to be a good strategy to design quantum spherical codes~\cite{Jain2024}, suggesting that such an approach would be fruitful in the context of quantum error correction.

\section{Discussion}

We introduced~\fashion, a fast optimization algorithm to enforce pair correlations in point patterns on the sphere through their spectral representation.
Having established its quasilinear scaling against the number of points and the considerable improvement it represents compared to previously available methods, we demonstrated its capabilities on a variety of systems of interest, in particular hyperuniform ones.
This work paves the way for the systematic generation of optimal point patterns to sample functions on the sphere, with applications spanning computer science~\cite{Pilleboue2015,Pharr2018}, signal acquisition, and processing across fields of experimental and computational science~\cite{Potts2017,Chen2018a,Zhou2018a_fixed,Lin2024,Xiao2025,Rafaely2005,Pinardi2021,Tanaka2025,Marantis2009,Chao2014,Flyer2012,Ishioka2018}.

From a fundamental perspectives,~\fashion~makes it possible to systematically assess the realizability of point patterns with specific long-range correlation structures on the sphere, a problem that has so far largely been confined to flat space but has attracted a lot of interest in the random geometry literature~\cite{Yamada1961,Kuna2007,Lachieze-Rey2025}.
More concretely, assessing the number of patterns with specified correlations on the sphere is relevant to the fundamental understanding of dynamics that spontaneously organize into long-range correlations, for instance biophysical processes~\cite{Lorman2007,Radja2019}.
In particular, there has recently been interest for the optimal repartition of receptors at the surface of cells~\cite{Alonso2025,Wu2025}, a problem that could easily be reframed as an optimal sampling one.
Patterning on the sphere also relates to the problem of circle packing on spheres, which has been the study of a number of works in physics both in the crystal~\cite{Bowick2000,Irvine2010,Fantoni2012} and glassy~\cite{Vest2014,Vest2018} contexts.

Our optimization method is in no way limited to flat and spherical manifolds: in fact, provided that one may compute the eigenfunctions of the Laplace-Beltrami operator and define the associated Manifold Harmonic Transform (MHT), one may in principle follow the same algorithmic route on any curved surface.
A promising route is that of computation on tessellated surfaces~\cite{Lévy2006,Vallet2008} which, while they are costlier than NUFFT or NUSHT algorithms, would allow to design optimal sampling schemes on arbitrary shapes, or to design truly curvature-aware coatings for photonics.

Finally, NUSHT algorithms typically come with spinful spherical harmonics representations~\cite{Reinecke2013,Belkner2024}, so that one could extend~\fashion~to the space of spinful functions, with possible applications in cosmological signal processing.

\section{Acknowledgements}
M.C. thanks Aaron Shih, David Grier, Misaki Ozawa and Jorge Kurchan for insightful comments on this work.
The authors acknowledge financial support from the Simons Center for Computational Physical Chemistry and the AFOSR Young Investigator Program under award FA9550-25-1-0359.
This work was supported in part through the NYU IT High Performance Computing resources, services, and staff expertise.
Pair correlations, discrepancy measurements, and Voronoi tessellations used in farthest-first insertions are computed using the \texttt{rusted\_core}~\cite{rusted_core} package.

\bibliography{PostDoc-StefanoMartiniani,ref}

\appendix

\section{Conventions \label{app:conventions}}

In this paper, we define spherical harmonics using the normalization and phase conventions such that 
\begin{align}
    Y_{\ell}^m(\theta, \phi) \equiv \sqrt{\frac{2\ell +1}{4\pi} \frac{(\ell - m)!}{(\ell + m)!}} P_{\ell}^m(\cos \theta) e^{i m \phi} \label{eq:Spherical_Harmonics}
\end{align}
with $P_{\ell}^m$ the \textit{associated Legendre polynomials},
\begin{align}
    P_{\ell}^m(z) = \frac{(-1)^m}{2^\ell \ell!} \left(1-z^2\right)^{m/2} \frac{d^{\ell +m}}{dz^{\ell +m}} \left[\left(z^2 - 1 \right)^\ell\right]. \label{eq:Associated_Legendre}
\end{align}
Note that Ref.~\cite{Abramowitz} distinguishes two separate kinds of associated Legendre polynomials through the notation
\begin{align}
    P_{\ell,m}(z) = (-1)^m P_\ell^m(z),
\end{align}
which is the reason for adopting the index-and-exponent notations in this paper for clarity.

These definitions ensure that spherical harmonics are orthonormal by the Hermitian inner product
\begin{align}
    \int\limits_{S_2} d^2 \Omega Y_{\ell}^m(\Omega) {Y_{\ell'}^{m'}}^\star(\Omega) = \delta_{\ell \ell'} \delta_{mm'}.
\end{align}
Note that with this definition, one has
\begin{align}
    Y_\ell^{-m} = (-1)^m {Y_\ell^m}^\star.
\end{align}

Since the spherical harmonics representation verifies, $\forall f,g \in L^2(S_2\mapsto \mathbb{C})$,
\begin{align}
    \int\limits_{S_2} d^2\Omega f(\theta, \phi) g^\star(\theta,\phi) = \sum\limits_{\ell = 0}^\infty \sum\limits_{m=-\ell}^\ell \wideparen{f}_{\ell}^m \wideparen{g}_\ell^m{}^{\star},
\end{align}
one may then introduce a power spectrum for $\wideparen{\rho}$ as
\begin{align}
    S_{\ell}^m \equiv \frac{4\pi}{N}\left |\wideparen{\rho}_{\ell}^m \right|^2,
\end{align}
or an $\ell$-only version as
\begin{align}
    S_{\ell} \equiv \frac{4\pi}{N} \frac{1}{2\ell + 1} \sum\limits_{m=-\ell}^\ell\left |\wideparen{\rho}_{\ell}^m \right|^2.
\end{align}
In particular, noticing that $Y_0^0 = 1/\sqrt{4\pi}$, one finds 
\begin{align}
    S_0 = S_0^0 =  \frac{1}{ N} \sum\limits_{n=1}^N |c_n|^2.
\end{align}
In particular, for equally-weigthed points, $c_n = 1$, this yields $S_0 = N$.
This expression is thus analogous to the usual Fourier-space power spectrum of point patterns in Euclidean space, as used in Ref.~\cite{Shih2023}.

Note that if $\rho$ is real-valued, the spherical harmonics representation has a Friedel-like symmetry,
\begin{align}
    \wideparen{\rho}_\ell^{m} &= \int\limits_{S^2} d^2\Omega \,\rho(\theta, \phi) {Y_{\ell}^{m}}^\star(\theta, \phi) \nonumber\\
    &= (-1)^m \int\limits_{S^2} d^2\Omega \,\rho(\theta, \phi) {Y_{\ell}^{-m}}(\theta, \phi) \nonumber\\
    &= (-1)^m \wideparen{\rho}_\ell^{-m}{}^\star. \label{eq:realvalued_rho_SHT}
\end{align}
As a result, it is common, when assuming only real-valued fields, to compute only components with $m\geq 0$.

Furthermore, if $\rho$ is complex-valued, it is useful to note that
\begin{align}
    \text{Re}\left[ \rho(\theta, \phi) \right] &= \frac{1}{2}\left[ \rho(\theta, \phi) + \rho(\theta,\phi)^\star \vphantom{\sum} \right] \nonumber \\
    &=\frac{1}{2}\sum\limits_{\ell = 0}^\infty \sum\limits_{m= -\ell}^\ell \left[ \wideparen{\rho}_{\ell}^m Y_{\ell}^m(\theta, \phi) + \wideparen{\rho}_{\ell}^m{}^\star Y_{\ell}^m{}^\star(\theta, \phi)  \right] \nonumber\\
    &= \frac{1}{2}\sum\limits_{\ell = 0}^\infty \sum\limits_{m= -\ell}^\ell \left[ \wideparen{\rho}_{\ell}^m Y_{\ell}^m(\theta, \phi) + (-1)^m\wideparen{\rho}_{\ell}^m{}^\star Y_{\ell}^{-m}(\theta, \phi)  \right] \nonumber \\
    &= \sum\limits_{\ell = 0}^\infty \sum\limits_{m= -\ell}^\ell \frac{1}{2} \left[ \wideparen{\rho}_{\ell}^m  + (-1)^m\wideparen{\rho}_{\ell}^{-m}{}^\star   \right] Y_{\ell}^m(\theta, \phi). \label{eq:realpart_ISHT}
\end{align}
Likewise,
\begin{align}
    \text{Im}\left[ \rho(\theta, \phi) \right] &= \frac{1}{2i}\left[ \rho(\theta, \phi) - \rho(\theta,\phi)^\star \vphantom{\sum} \right] \nonumber \\
    &=\frac{1}{2i}\sum\limits_{\ell = 0}^\infty \sum\limits_{m= -\ell}^\ell \left[ \wideparen{\rho}_{\ell}^m Y_{\ell}^m(\theta, \phi) -\wideparen{\rho}_{\ell}^m{}^\star Y_{\ell}^m{}^\star(\theta, \phi)  \right] \nonumber \\
    &= \frac{1}{2i}\sum\limits_{\ell = 0}^\infty \sum\limits_{m= -\ell}^\ell \left[ \wideparen{\rho}_{\ell}^m Y_{\ell}^m(\theta, \phi) - (-1)^m\wideparen{\rho}_{\ell}^m{}^\star Y_{\ell}^{-m}(\theta, \phi)  \right] \nonumber \\
    &= \sum\limits_{\ell = 0}^\infty \sum\limits_{m= -\ell}^\ell \frac{1}{2i} \left[ \wideparen{\rho}_{\ell}^m  + (-1)^{m+1}\wideparen{\rho}_{\ell}^{-m}{}^\star   \right] Y_{\ell}^m(\theta, \phi), \nonumber \\
    &= \sum\limits_{\ell = 0}^\infty \sum\limits_{m= -\ell}^\ell \frac{i}{2} \left[ -\wideparen{\rho}_{\ell}^m  + (-1)^{m}\wideparen{\rho}_{\ell}^{-m}{}^\star   \right] Y_{\ell}^m(\theta, \phi). \label{eq:imagpart_ISHT}
\end{align}

\section{Gradient of the loss\label{app:Gradients}}

One may write the gradients of the loss, Eq.~\eqref{eq:Loss}, with respect to a coordinate $x_p$ that may either be a position coordinate, $\theta_p$ or $\phi_p$, or a coefficient $c_p$ for point $p$,
\begin{align}
    \frac{\partial L}{\partial x_p}[S,S_0] = 2\sum\limits_{\ell\in \Lambda} W_{\ell} \frac{\partial S_\ell}{\partial x_p}\left[ S_\ell - S_{0}(\ell) \right]. \label{eq:Loss_gradient_generic}
\end{align}
The derivative of the power spectrum component $S_\ell$ can be expressed as
\begin{align}
    \frac{\partial S_\ell}{\partial x_p} = \frac{8\pi}{(2\ell + 1)N} \sum\limits_{m= - \ell}^\ell \text{Re}\left[ \frac{\partial \wideparen{\rho}_{\ell}^m}{\partial x_p } \wideparen{\rho}_{\ell}^m{}^\star\right]. \label{eq:Sell_gradient_generic}
\end{align}
Recalling the expression of $\wideparen{\rho}$, Eq.~\eqref{eq:rho_SHT}, one furthermore has
\begin{align}
    \frac{\partial \wideparen{\rho}_{\ell}^m}{\partial c_p} &= Y_{\ell}^m{}^\star (\theta_p, \phi_p) \label{eq:rhograd_c} \\
    \frac{\partial \wideparen{\rho}_{\ell}^m}{\partial \theta_p} &= c_p \partial_\theta Y_{\ell}^m{}^\star (\theta_p, \phi_p) \label{eq:rhograd_theta}\\
    \frac{\partial \wideparen{\rho}_{\ell}^m}{\partial \phi_p} &= c_p \partial_\phi Y_{\ell}^m{}^\star (\theta_p, \phi_p) . \label{eq:rhograd_phi}
\end{align}
Finally, using the definition of spherical harmonics, Eq.~\eqref{eq:Spherical_Harmonics},
\begin{align}
    \partial_\phi Y_{\ell}^m(\theta, \phi) &= i m Y_{\ell}^m(\theta, \phi) \label{eq:Ylm_gradient_phi}
\end{align}
and
\begin{align}
    &\partial_\theta Y_{\ell}^m(\theta, \phi) = \sqrt{\frac{2\ell +1}{4\pi} \frac{(\ell - m)!}{(\ell + m)!}} \partial_\theta P_{\ell}^m(\cos \theta) e^{i m \phi} \nonumber \\
    &= -\sin\theta \sqrt{\frac{2\ell +1}{4\pi} \frac{(\ell - m)!}{(\ell + m)!}}  P_{\ell}^m{}'(\cos \theta) e^{i m \phi} \nonumber \\
    &= -\sqrt{\frac{2\ell +1}{4\pi} \frac{(\ell - m)!}{(\ell + m)!}}  \sqrt{1 -\cos^2\theta}P_{\ell}^m{}'(\cos \theta) e^{i m \phi} \label{eq:Ylm_partialtheta_step1}
\end{align}
with, using Eq.~\eqref{eq:Associated_Legendre}, for $m< \ell$,
\begin{widetext}
\begin{align}
    -\sqrt{1-z^2}P_{\ell}^m{}'(z) &= \frac{(-1)^m}{2^\ell \ell!} (1-z^2)^{1/2}\left(  mz (1-z^2)^{m/2-1} \frac{d^{\ell +m}}{dz^{\ell +m}} \left[\left(z^2 - 1 \right)^\ell\right] - (1-z^2)^{m/2} \frac{d^{\ell +m+1}}{dz^{\ell +m+1}} \left[\left(z^2 - 1 \right)^\ell\right] \right) \nonumber \\
    &= \frac{(-1)^m}{2^\ell \ell!} \left(  mz (1-z^2)^{(m-1)/2} \frac{d^{\ell +m}}{dz^{\ell +m}} \left[\left(z^2 - 1 \right)^\ell\right] - (1-z^2)^{(m+1)/2} \frac{d^{\ell +m+1}}{dz^{\ell +m+1}} \left[\left(z^2 - 1 \right)^\ell\right] \right) \nonumber  \\
    &= \frac{m z}{\sqrt{1-z^2}} P_\ell^m(z) + P_\ell^{m+1}(z)
\end{align}
\end{widetext}
and, for $m = \ell$, since $\text{deg}[P_{\ell}^m] = \ell -m$,
\begin{align}
    -\sqrt{1-z^2}P_{\ell}^\ell{}'(z) &= 0.
\end{align}
Realizing that
\begin{align}
    \frac{\cos\theta}{\sqrt{1-\cos^2\theta}} = \cot\theta,
\end{align}
this expression can be injected into Eq.~\eqref{eq:Ylm_partialtheta_step1}, one finds
\begin{widetext}
\begin{align}
    \partial_\theta Y_{\ell}^m(\theta, \phi) &= \sqrt{\frac{2\ell +1}{4\pi} \frac{(\ell - m)!}{(\ell + m)!}} \left[m\cot \theta P_\ell^m(\cos\theta) + P_\ell^{m+1}(\cos\theta)\right] e^{i m \phi} \nonumber \\
    &= m\cot\theta Y_{\ell}^m(\theta, \phi) + \sqrt{(\ell-m)(\ell+m+1)}e^{-i\phi}Y_{\ell}^{m+1}(\theta, \phi). \label{eq:Ylm_gradient_theta}
\end{align}
\end{widetext}

All in all, injecting the complex conjugates of the components of the gradient of spherical harmonics, Eqs.~\eqref{eq:Ylm_gradient_theta} and~\eqref{eq:Ylm_gradient_phi} into the gradients of $\wideparen{\rho}$, Eqs.~\eqref{eq:rhograd_c},~\eqref{eq:rhograd_theta} and~\eqref{eq:rhograd_phi} yields
\begin{widetext}
\begin{align}
    \frac{\partial \wideparen{\rho}_{\ell}^m}{\partial c_p} &= Y_{\ell}^m{}^\star (\theta_p, \phi_p) \label{eq:rhograd_c_explicit} \\
    \frac{\partial \wideparen{\rho}_{\ell}^m}{\partial \theta_p} &= c_p \left( m\cot\theta Y_{\ell}^m{}^\star(\theta, \phi) + \sqrt{(\ell-m)(\ell+m+1)}e^{i\phi}Y_{\ell}^{m+1}{}^\star(\theta, \phi). \right) \label{eq:rhograd_theta_explicit}\\
    \frac{\partial \wideparen{\rho}_{\ell}^m}{\partial \phi_p} &= - i m c_p Y_{\ell}^m{}^\star (\theta_p, \phi_p) . \label{eq:rhograd_phi_explicit}
\end{align}
\end{widetext}
These equations can then be injected into the components of the gradient of the power spectrum, Eq.~\eqref{eq:Sell_gradient_generic}, to yield
\begin{widetext}
\begin{align}
    \frac{\partial S_\ell}{\partial c_p} &= \frac{8\pi}{(2\ell + 1)N} \sum\limits_{m= - \ell}^\ell \text{Re}\left[  Y_{\ell}^m{}^\star (\theta_p, \phi_p)\wideparen{\rho}_{\ell}^m{}^\star\right], \\
    \frac{\partial S_\ell}{\partial \theta_p} &= \frac{8\pi}{(2\ell + 1)N} \sum\limits_{m= - \ell}^\ell \text{Re}\left[ c_p \left( m\cot\theta Y_{\ell}^m{}^\star(\theta, \phi) + \sqrt{(\ell-m)(\ell+m+1)}e^{i\phi}Y_{\ell}^{m+1}{}^\star(\theta, \phi). \right)\wideparen{\rho}_{\ell}^m{}^\star\right], \\
    \frac{\partial S_\ell}{\partial \phi_p} &= \frac{8\pi}{(2\ell + 1)N} \sum\limits_{m= - \ell}^\ell \text{Re}\left[ - i m c_p Y_{\ell}^m{}^\star (\theta_p, \phi_p)\wideparen{\rho}_{\ell}^m{}^\star\right].
\end{align}
\end{widetext}

Finally, these expressions can be injected into Eq.~\eqref{eq:Loss_gradient_generic}, to yield the gradients of the loss, which may be written as inverse Spherical Harmonics Transforms.
First, the gradient with respect to coefficients reads
\begin{align}
    \frac{\partial L}{\partial c_p} &= \text{Re}\left[\sum\limits_{\ell \in \Lambda} \sum\limits_{m= - \ell}^\ell \wideparen{C}_{\ell}^m Y_{\ell}^m{} (\theta_p, \phi_p)\right] \label{eq:Loss_gradient_c} \\
    \wideparen{C}_\ell^m &\equiv  \frac{16\pi}{(2\ell + 1)N}   W_\ell (S_\ell - S_0(\ell)) \wideparen{\rho}_{\ell}^m{} \label{eq:SHT_cp_grad}
\end{align}
where we assumed that $W_\ell \in \mathbb{R}$ and used the fact that $\text{Re}[z] = \text{Re}[z^\star]$.
Likewise, the gradient with respect to the azimuthal component reads
\begin{align}
    \frac{\partial L}{\partial \phi_p} &= \text{Re}\left[\sum\limits_{\ell \in \Lambda} \sum\limits_{m= - \ell}^\ell \wideparen{D}_{\ell}^m Y_{\ell}^m{} (\theta_p, \phi_p)\right] \label{eq:Loss_gradient_phi} \\
    \wideparen{D}_\ell^m &\equiv \frac{16 i m\pi c_p^\star}{(2\ell + 1)N}   W_\ell (S_\ell - S_0(\ell)) \wideparen{\rho}_{\ell}^m{}.
\end{align}
Finally, the gradient with respect to $\theta_p$ can be written as
\begin{widetext}
\begin{align}
    \frac{\partial L}{\partial \theta_p} &= \text{Re}\left[\sum\limits_{\ell \in \Lambda} \sum\limits_{m= - \ell}^\ell \wideparen{E}_{\ell}^m Y_{\ell}^m{} (\theta_p, \phi_p)\right]\label{eq:Loss_gradient_theta} \\
    \wideparen{E}_\ell^m &\equiv \frac{16 \pi c_p^\star  }{(2\ell + 1)N}   W_\ell (S_\ell - S_0(\ell)) \left[ m\cot \theta_p\wideparen{\rho}_{\ell}^m + \sqrt{(\ell - m + 1)(\ell + m)}e^{-i\phi_p}\wideparen{\rho}_{\ell}^{m-1}{}\right].
\end{align}
\end{widetext}
To obtain the last expression, one needs to notice that
\begin{widetext}
\begin{align}
    \sigma_\ell(\theta, \phi) &\equiv \sum\limits_{m=-\ell}^\ell \sqrt{(\ell-m)(\ell+m+1)}Y_{\ell}^{m+1}\wideparen{\rho}_{\ell}^m \nonumber \\
    &= \sum\limits_{m=-\ell +1}^{\ell+1} \sqrt{(\ell - m +1)(\ell +m)} Y_{\ell}^m(\theta, \phi) \wideparen{\rho}_\ell^{m-1} &\text{( m $\mapsto$ m+1 )} \nonumber \\
    &= \sum\limits_{m=-\ell +1}^{\ell} \sqrt{(\ell - m +1)(\ell +m)} Y_{\ell}^m(\theta, \phi) \wideparen{\rho}_\ell^{m-1} &\text{( $ m = \ell+1$ term is zero )} \nonumber \\
    &= \sum\limits_{m=-\ell}^{\ell} \sqrt{(\ell - m +1)(\ell +m)} Y_{\ell}^m(\theta, \phi) \wideparen{\rho}_\ell^{m-1} &\text{( $ m = -\ell$ term is zero )}
\end{align}
\end{widetext}

The loss can thus be computed from $S_\ell$, which may itself be computed using a non-uniform fast spherical harmonics transform (NUFSHT), while its loss can be computed using the value of $\wideparen{\rho}_\ell$ as well as a small number of inverse NUFSHTs.
Note that the gradient with respect to $\theta_p$ diverges in the limit $\theta_p \to 0$, as the gradient is ill-defined at poles.

In practice, libraries that compute SHTs commonly assume that the field is originally real-valued to compute and store only SHT coefficients with $m \geq 0$.
It is thus convenient to assume $\rho \in \mathbb{R}$ and to use Eq.~\eqref{eq:realpart_ISHT} to write explicitly
\begin{align}
    \frac{\partial L}{\partial \phi_p} &= \sum\limits_{\ell \in \Lambda} \sum\limits_{m=-\ell}^\ell \wideparen{\mathcal{D}}_{\ell,p}^m Y_{\ell}^m (\theta_p, \phi_p) , \\
    \frac{\partial L}{\partial \theta_p} &= \sum\limits_{\ell \in \Lambda} \sum\limits_{m=-\ell}^\ell \wideparen{\mathcal{E}}_{\ell,p}^m Y_{\ell}^m (\theta_p, \phi_p) ,
\end{align}
with
\begin{align}
    \wideparen{\mathcal{D}}_{\ell,p}^m &\equiv \frac{1}{2} \left[ \wideparen{D}_\ell^m + (-1)^m \wideparen{D}_\ell^{-m}{}^\star\right] \nonumber \\
    &= \frac{8 \pi c_p}{(2\ell + 1)N}   W_\ell (S_\ell - S_0(\ell)) i m \left[\wideparen{\rho}_{\ell}^m{} +(-1)^m \wideparen{\rho}_{\ell}^{-m}{}^\star \right] \nonumber \\
    &= \frac{16 \pi c_p}{(2\ell + 1)N}   W_\ell (S_\ell - S_0(\ell)) i m \wideparen{\rho}_{\ell}^m{}
\end{align}
where in the last line I used Eq.~\eqref{eq:realvalued_rho_SHT}.
In practice, it is convenient to factor out all factors that depend on index $p$ or that are constant with respect to $\ell,m$ (save for the factor of $i$ that ensures real-valued quantities), yielding
\begin{align}
    \frac{\partial L}{\partial \phi_p} &= \frac{16\pi c_p}{N}\sum\limits_{\ell \in \Lambda} \sum\limits_{m=-\ell}^\ell \wideparen{\mathcal{D}}_\ell^m Y_{\ell}^m (\theta_p, \phi_p) \\
    \wideparen{\mathcal{D}}_\ell^m &\equiv i m\frac{W_\ell (S_\ell - S_0(\ell))}{2\ell+1} \wideparen{\rho}_{\ell}^m{}.
\end{align}

Likewise,
\begin{widetext}
\begin{align}
    \wideparen{\mathcal{E}}_{\ell,p}^m &\equiv \frac{1}{2} \left[ \wideparen{E}_\ell^m + (-1)^m \wideparen{E}_\ell^{-m}{}^\star\right] \nonumber \\
    &= \frac{8 \pi c_p  }{(2\ell + 1)N}   W_\ell (S_\ell - S_0(\ell)) \cot \theta_p m\left[  \wideparen{\rho}_{\ell}^m  -(-1)^m \wideparen{\rho}_{\ell}^{-m}{}^\star \right] \nonumber \\
    &+ \frac{8 \pi c_p  }{(2\ell + 1)N}   W_\ell (S_\ell - S_0(\ell)) \left[  \sqrt{(\ell - m + 1)(\ell + m)}e^{-i\phi_p}\wideparen{\rho}_{\ell}^{m-1}{} + (-1)^m \sqrt{(\ell + m + 1)(\ell - m)}e^{+i\phi_p}\wideparen{\rho}_{\ell}^{-m-1}{}^\star\right] \nonumber \\
    &= \frac{8 \pi c_p  }{(2\ell + 1)N}   W_\ell (S_\ell - S_0(\ell)) \left[  \sqrt{(\ell - m + 1)(\ell + m)}e^{-i\phi_p}\wideparen{\rho}_{\ell}^{m-1}{} - \sqrt{(\ell + m + 1)(\ell - m)}e^{+i\phi_p}\wideparen{\rho}_{\ell}^{m+1}{}\right]
\end{align}
\end{widetext}
where we used that Eq.~\eqref{eq:realvalued_rho_SHT} cancels out the first contribution -- which is reassuring as it removes the possible divergence if $\cot \theta$ at poles.
One then only needs to specify $\wideparen{\mathcal{D}}_\ell^m$ and $\wideparen{\mathcal{E}}_\ell^m$ for $m \geq 0$.
The only special edge case is that of $\wideparen{\mathcal{E}}_\ell^0$, which reads
\begin{align}
    \wideparen{\mathcal{E}}_\ell^0 =&\frac{8 \pi c_p  }{(2\ell + 1)N}   W_\ell (S_\ell - S_0(\ell)) \sqrt{\ell(\ell + 1)} \nonumber \\
    &\qquad \qquad \qquad  \times \left[  e^{-i\phi_p}\wideparen{\rho}_{\ell}^{-1}{} -e^{+i\phi_p}\wideparen{\rho}_{\ell}^{+1}{}\right].
\end{align}
Per Eq.~\eqref{eq:realvalued_rho_SHT}, 
\begin{align}
    \wideparen{\rho}_{\ell}^{-1}{} = - \wideparen{\rho}_{\ell}^{+1}{}^\star
\end{align}
so that
\begin{align}
    \wideparen{\mathcal{E}}_{\ell,p}^0 &=\frac{8 \pi c_p  }{(2\ell + 1)N}   W_\ell (S_\ell - S_0(\ell)) \sqrt{\ell(\ell + 1)} \nonumber \\
    &\qquad \qquad\qquad  \times \left[ e^{+i\phi_p}\wideparen{\rho}_{\ell}^{+1}{} +  \left(e^{+i\phi_p}\wideparen{\rho}_{\ell}^{+1}{}\right)^\star\right] \nonumber \\
    &= \frac{16 \pi c_p  }{(2\ell + 1)N}   W_\ell (S_\ell - S_0(\ell)) \sqrt{\ell(\ell + 1)} \,\text{Re}\left[ e^{+i\phi_p}\wideparen{\rho}_{\ell}^{+1}{}\right].
\end{align}
The other edge case, $m = \ell$, requires less effort to write explicitly, as it immediately reads
\begin{align}
    \wideparen{\mathcal{E}}_{\ell,p}^\ell &= \frac{8 \pi c_p  }{(2\ell + 1)N}   W_\ell (S_\ell - S_0(\ell)) \left[  \sqrt{2\ell }e^{-i\phi_p}\wideparen{\rho}_{\ell}^{\ell-1}{} \right].
\end{align}
In practice, it is convenient in code to factor out all functions of $p$ again, which here leads to a partition of the gradient into two separate transforms:
\begin{widetext}
\begin{align}
    \frac{\partial L}{\partial \theta_p} &= \frac{8\pi c_p}{N} \cos\phi_p\sum\limits_{\ell \in \Lambda} \sum\limits_{m=-\ell}^\ell \wideparen{\mathcal{F}}_{\ell}^m Y_{\ell}^m (\theta_p, \phi_p) + \frac{8\pi c_p}{N} \sin\phi_p\sum\limits_{\ell \in \Lambda} \sum\limits_{m=-\ell}^\ell \wideparen{\mathcal{G}}_{\ell}^m Y_{\ell}^m (\theta_p, \phi_p), \\
    \wideparen{\mathcal{F}}_{\ell}^m &\equiv \frac{W_\ell (S_\ell - S_0(\ell))}{2\ell+1} \left[  \sqrt{(\ell -m+1)(\ell+m)}\wideparen{\rho}_{\ell}^{m-1}{} - \sqrt{(\ell +m+1)(\ell-m)}\wideparen{\rho}_{\ell}^{m+1}{}\right], \\
    \wideparen{\mathcal{G}}_{\ell}^m &\equiv -i \frac{W_\ell (S_\ell - S_0(\ell))}{2\ell+1} \left[  \sqrt{(\ell -m+1)(\ell+m)}\wideparen{\rho}_{\ell}^{m-1}{} + \sqrt{(\ell +m+1)(\ell-m)}\wideparen{\rho}_{\ell}^{m+1}{} \right].
\end{align}  
\end{widetext}

Note that the gradient with respect to $c_p$, Eq.~\eqref{eq:SHT_cp_grad}, is already real-valued as long as $\rho$ itself is real-valued.

\section{Optimization method \label{app:Optimization}}

We here provide more details on the optimization method.

\subsection{Tangent-space evolution and projection}

We first remind the basics of how the spherical gradient is used to propose an update in the tangent space then project.
The key is to write positions $(\theta, \phi)$ on the sphere as positions in the embedding space, here $\mathbb{R}^3$,
\begin{align}
    \bm{r} = \hat{\bm{e}}_r(\theta, \phi)=\sin\theta\cos\phi \hat{\bm{e}}_x + \sin\theta\sin\phi \hat{\bm{e}}_y + \cos\phi \hat{\bm{e}}_z
\end{align}
and to define the other two vectors that form the conventional basis of the spherical coordinate system,
\begin{align}
    \hat{\bm{e}}_\theta & =  \cos\theta\cos\phi\hat{\bm{e}}_x +  \cos\theta \sin\phi \hat{\bm{e}}_y -\sin\theta \hat{\bm{e}}_z, \\
    \hat{\bm{e}}_\phi &= -\sin\phi \hat{\bm{e}}_x + \cos\phi  \hat{\bm{e}}_y.
\end{align}

With these conventions, one may write the gradient of the analytic continuation of the loss function to $\mathbb{R}^3$ as
\begin{align}
    \bm{\nabla}L &= \sum\limits_{n=1}^N\left[ \frac{\partial L}{\partial \theta_n} \bm{\hat{e}_{\theta_n}} +  \frac{1}{\sin\theta_n}\frac{\partial L}{\partial \phi_n} \bm{\hat{e}_{\phi_n}}\right] \\
    &=\sum\limits_{n=1}^N \left[ \frac{\partial L}{\partial x_{n}} \bm{\hat{e}}_{x} + \frac{\partial L}{\partial y_{n}} \bm{\hat{e}}_{y} + \frac{\partial L}{\partial z_{n}} \bm{\hat{e}}_{z} \right].
\end{align}
Using this last identity, one may update the position of point $n$ using the Cartesian representation of the gradient within the chosen gradient-based optimization scheme.
After this update, the point $\bm{r}_n(t+\Delta t)$ generically lies outside of the sphere, we thus project it back and compute $\theta_n(t+\Delta t) = \arccos [z_n(t+\Delta t) / r_n(t+\Delta t)]$ and $\phi_n(t+\Delta t) = \arctan [y_n(t+\Delta t) / x_n(t+\Delta t)]$.

\subsection{Optimizer parameters}

Throughout this work, we use the same L-BFGS implementation that was used in FReSCo~\cite{FReSCo} or in Ref.~\cite{Suryadevara2024a}.
The specificity of this implementation is that we enforce a maximum step size $\Delta x_{\max} = 0.01$ radians.
Termination is reached either when the root-mean-square value of the gradient vector (in Cartesian representation) is lower than $g = 10^{-30}$, or when the number of steps reaches $n_{max} = 10^5$.
The number of steps used by L-BFGS to construct an estimate of the Hessian is set to $M = 20$.

\subsection{Gyromorph generation strategy}

Like in flat space~\cite{Casiulis2024a}, maximizing the height of a mode at a specific $\ell$ value without any other constraint generically leads to exact point overlaps.
Adding pair repulsion between points however creates a more rugged loss landscape and may create non-trivial undesirable configurations as the output of optimization.
We thus adopt the same strategy as in Ref.~\cite{Casiulis2024a} -- namely, we cycle 2 phases.
The first one is standard~\fashion~optimization.
In the second one, we find all points that have neighbors below a threshold distance smaller than the mean-field distance between points $d_{ov} \ll \pi$, thus identifying clusters of near-overlapping points.
In each of these clusters, we only retain one point.
We then re-inject the $n$ removed points to keep the total number of points $N$ constant.
To re-inject them, we compute the Voronoi tessellation of the remaining points, and find the $n$ vertices that lie farthest from neighboring points, where we place the points.
In this paper, we consistently use $d_{ov} = 0.7\pi /\ell_{\max}$, where $\pi / \ell_{\max}$ is used as a proxy for the typical distance between points.

\section{Centrosymmetric patterns \label{app:CentroLoss}}

A special case used in the main text is that of centrosymmetric point patterns on the sphere, namely patterns such that $N = 2 M$ and $\forall m \in \{1,\ldots,M\}$, $\bm{r}_m = - \bm{r}_{m+M}$ and $c_m = c_{m+M}$.
In spherical coordinates, this translates into 
\begin{align}
    \theta_{m+M} &= \pi - \theta_m, \\
    \phi_{m+M} &\equiv \phi_m + \pi \mod 2\pi.
\end{align}
Using the definition of spherical harmonics, this implies
\begin{align}
    \wideparen{\rho}_{\ell}^m &= \sum\limits_{n=1}^{2M} c_n Y_{\ell}^m{}^\star(\theta_n, \phi_n) \nonumber \\
    &= \sum\limits_{m=1}^{M} c_m \left( Y_{\ell}^m{}^\star(\theta_n, \phi_n) + Y_{\ell}^m{}^\star(\pi - \theta_n, \phi_n + \pi) \right) \nonumber \\
    &= \sum\limits_{n=1}^{M} c_n Y_{\ell}^m{}^\star(\theta_n, \phi_n) \left[1 + (-1)^\ell \right] \nonumber  \\
    &= \begin{cases}
        2 \wideparen{\rho}_{\ell,1/2}^m & \text{if } \ell \in 2\mathbb{N} \\
        0 & \text{otherwise}
    \end{cases}
\end{align}
where we defined
\begin{align}
    \wideparen{\rho}_{\ell,1/2}^m \equiv \sum\limits_{n=1}^{M} c_n Y_{\ell}^m{}^\star(\theta_n, \phi_n)
\end{align}
the SHT of an irreducible half of the points.

Thus, the problem of optimizing such point patterns is essentially unchanged, up to a factor of $2$ in the value of the maximal $\ell$ that can be constrained.

\section{Pair correlations on the surface of a sphere\label{app:PairCorr}}

We now briefly discuss the notion of a pair correlation function, usually called $g(\bm{r})$, on the sphere.

\subsection{Generic considerations on $g$}

The most common setting to discuss correlations in point patterns is that of \textit{random} point patterns, in which the whole pattern is drawn from a joint probability distribution function with measure $\mu(\bm{r}_1, \bm{r}_2, \ldots,\bm{r}_N)$ so that, in particular
\begin{align}
    \left\langle \rho(\bm{r})\right\rangle = \int d^d\bm{r}_1 \ldots d^d\bm{r}_N \mu(\bm{r}_1, \ldots ,\bm{r}_N) \rho(\bm{r}).
\end{align}
In the simple example of a Poisson point pattern, the measure is simply Lebesgue over the volume $V$ of the domain in which points are drawn, so that
\begin{align}
    \mu(\bm{r}_1, \ldots, \bm{r}_N) = \frac{1}{V^N}
\end{align}
and 
\begin{align}
    \left\langle \rho(\bm{r})\right\rangle = \frac{N}{V} \overline{c_n} \equiv \rho_0,
\end{align}
where we introduced the arithmetic mean of coefficients,
\begin{align}
    \overline{c_n} = \frac{1}{N} \sum\limits_{n=1}^N c_n.
\end{align}
One often considers the broader family of translation-invariant (also called homogeneous or stationary) point patterns, such that $\langle \rho (\bm{r})\rangle = \rho_0$ but $\mu \neq 1/V$ in such a way that there are correlations between point positions.
The observables of interest in that situation typically have to do with the correlations between the positions of points.
The most common object to study in physics is the pair correlation function (PCF) (also called pair distribution function or PDF in the literature, which we here avoid to avoid ambiguities with probability density function), $g(\bm{r}, \bm{r}')$, which may be defined through~\cite{Hansen2006}
\begin{align}
    g(\bm{r}, \bm{r}') \equiv \frac{\left\langle \rho(\bm{r})  \rho(\bm{r'})\right\rangle}{\left\langle \rho(\bm{r}) \right\rangle \left\langle  \rho(\bm{r'})\right\rangle}. \label{eq:general_g2_nocenterremoval}
\end{align}
It is common to define a slight variation of it that excludes the case $\bm{r} = \bm{r'}$ explicitly~\cite{Hansen2006},
\begin{align}
    g(\bm{r}, \bm{r}') \equiv \frac{\left\langle \rho(\bm{r})  \rho(\bm{r'})\right\rangle - \langle \rho (\bm{r}) \rangle \delta(\bm{r} - \bm{r}')}{\left\langle \rho(\bm{r}) \right\rangle \left\langle  \rho(\bm{r'})\right\rangle}.
\end{align}
For a homogeneous point pattern, this last definition simplifies to
\begin{align}
    g(\bm{r}, \bm{r}') \equiv \frac{\left\langle \rho(\bm{r})  \rho(\bm{r'})\right\rangle - \rho_0 \delta(\bm{r} - \bm{r}')}{\rho_0^2}. \label{eq:general_g2_stats}
\end{align}
Furthermore, assuming translational invariance means that both measurement points may be translated by $\bm{r}$, so that
\begin{align}
    g(\bm{r}, \bm{r}') = g(\bm{r}-\bm{r}').
\end{align}

When given a single point pattern, one often defines the pair correlation function of the density field as
\begin{align}
g(\bm{r}) \equiv V \frac{\int d^d \bm{r}_1 d^d \bm{r}_2 \rho(\bm{r}_1) \rho(\bm{r}_2) \delta(\bm{r} - \bm{r}_{12})}{\left(\int d^d \bm{r}_1 \rho(\bm{r}_1)\right) \left( \int d^d \bm{r}_2 \rho(\bm{r}_2)\right)},
\end{align}
which tends to $1$ as the density fields at positions $\bm{r}_1$ and $\bm{r}_1 + \bm{r}$ become independent, usually as $r \to \infty$.
This expression can also be simplified, leading to
\begin{align}
g(\bm{r}) &= V \frac{\int d^d \bm{r}_1 \rho(\bm{r}_1) \rho(\bm{r}_1 + \bm{r})}{\left(\int d^d \bm{r}_1 \rho(\bm{r}_1)\right) \left( \int d^d \bm{r}_2 \rho(\bm{r}_2)\right)}, \\
          &= \frac{1}{\rho_0^2 V} \int d^d \bm{r}_1 \rho(\bm{r}_1) \rho(\bm{r}_1 + \bm{r}) \\
          &= \frac{V}{N^2 \overline{c_n}^2} \sum_{i,j} c_i c_j \delta(\bm{r} - \bm{r}_{ij}).
\end{align}
In practice, one often evaluates $g$ from discrete bins,
\begin{align}
    g_{\Delta r}(\bm{r}) &= \frac{1}{\rho_0 N \Delta r^2}\sum\limits_{i\neq j} \widehat{\delta}_{\Delta r}(\bm{r} - \bm{r}_{ij})
\end{align}
where $\widehat{\delta}$ is a binning function that generates a count histogram with square pixels, and $\Delta r$ is its bin size.

Furthermore, one often defines the radial correlation function (RDF).
In Euclidean space, $d^d\bm{r} = r^{d-1} dr d^{d-1}\Omega$ and one may define
\begin{align}
    g(r) \equiv \frac{1}{\Omega_d} \int d^{d-1}\Omega \,g(\bm{r})
\end{align}
with
\begin{align}
    \Omega_d \equiv \int d^{d-1}\Omega.
\end{align}
One then has 
\begin{align}
    \int_{\mathcal{R}} d^d \bm{r} g(\bm{r}) = \int dr f(r) r^{d-1} \Omega_d g(r) 
\end{align}
where $f(r)$ is a box-dependent function that represents the fraction of the hypersphere contained in the box at radius $r$.
In free space, $f(r) = 1$.
The histogrammed version of the RDF is then
\begin{align}
    g_{\Delta r}(r) &= \frac{1}{f(r) r^{d-1} \, \Delta r \, \Omega_d } 
 \frac{1}{\rho_0 N}\sum\limits_{i\neq j} \widehat{\delta}_{\Delta r}(r - r_{ij}).
\end{align}

Note in particular that, in a Poisson point pattern, in all bins,
\begin{align}
    \sum\limits_{i=1}^{N} \sum\limits_{j \neq i} \widehat{\delta}_{\Delta r}(\bm{r} - \bm{r}_{ij}) = \sum\limits_{i=1}^{N} \rho_0 \Delta r^2 = N \rho_0 \Delta r^2
\end{align}
so that $g_{\Delta r}(\bm{r}) = 1$ as expected, and the inverse Jacobian in the definition of the radial version also ensures  $g_{\Delta r}(r) = 1$.

\subsection{Special case of the sphere}

In the case of the sphere, the distance between two points $\bm{r}_1$ and $\bm{r}_2$ is given by an arclength on a great circle.
Assuming a sphere with radius $R$, this arclength is simply proportional to the usual arccos angular distance,
\begin{align}
    r_{12} = R \arccos \frac{\bm{r}_1 \cdot \bm{r}_2}{r_1 r_2}.
\end{align}
Parametrizing the positions using spherical coordinates, $\bm{r}_{1,2} = (\theta_{1,2}, \phi_{1,2})$, the dot-product may be rewritten as
\begin{align}
    \frac{\bm{r}_1 \cdot \bm{r}_2}{r_1 r_2} = \cos\theta_1 \cos\theta_2 + \sin\theta_1 \sin\theta_2 \cos (\phi_1 - \phi_2)
\end{align}
so that 
\begin{align}
    r_{12} = R \arccos\left[ \cos\theta_1 \cos\theta_2 + \sin\theta_1 \sin\theta_2 \cos (\phi_1 - \phi_2) \vphantom{\int}\right].
\end{align}
The definitions of $g(\bm{r})$ given in the previous subsection are still valid with this definition of distance, nothing that $V = 4\pi R^2$.
We henceforth set $R = 1$.

The equivalent of the radial distribution function (RDF), however, must be adapted to account for the non-Euclidean nature of spheres.
On the sphere, adopting standard notations for solid angles,
\begin{align}
    \int\limits_{{S}_2} d^2\Omega g(\Omega) =\int\limits_{-\pi}^\pi d\phi \int\limits_{0}^\pi d\theta \sin\theta g(\theta, \phi).
\end{align}
In this writing of the integral, one implicitly assumes that the pole $\theta = 0$ of the coordinates system is the reference point that distances are computed from.
The great-circle distance is then purely encoded by $\theta$, which plays the role of the radial coordinate in Euclidean space.
One may then introduce an RDF as
\begin{align}
    g(\theta) \equiv \frac{1}{2\pi} \int\limits_{-\pi}^\pi d\phi \, g(\theta, \phi),
\end{align}
so that
\begin{align}
    \int\limits_{{S}_2} d^2\Omega g(\Omega) &= 2\pi \int\limits_{0}^\pi d\theta \sin\theta g(\theta) \\
    &= 4\pi \int\limits_{0}^\pi d\theta f(\theta) g(\theta)
\end{align}
where 
\begin{align}
f(\theta) = \sin\theta/2
\end{align}
plays the same role as $f(r)$ in the Euclidean case: it is the ``radial density of mass'' of the sphere.
Thus, one may compute a binned version of the RDF on the sphere as
\begin{align}
    g_{\Delta r}(r) \equiv  \frac{1}{4\pi f(r) \, \Delta r \,  } 
 \frac{1}{\rho_0 N}\sum\limits_{i\neq j} \widehat{\delta}_{\Delta r}(r - r_{ij}),
\end{align}
or, counting each pair only once and replacing $f$ by its explicit expression,
\begin{align}
    g_{\Delta r}(r) &= \frac{1}{\pi \sin(r) \, \Delta r \,  } 
 \frac{1}{\rho_0 N}\sum\limits_{i< j} \widehat{\delta}_{\Delta r}(r - r_{ij}), \\
 &= \frac{4}{ N^2\sin(r) \, \Delta r \,  } 
 \sum\limits_{i< j} \widehat{\delta}_{\Delta r}(r - r_{ij}),
\end{align}

\subsection{Poisson point pattern on the sphere}

It is useful to consider uncorrelated, or ``Poisson'' point patterns on the sphere both as initial conditions and as analytically tractable examples of point patterns.
They are defined as random point patterns that have Lebesgue measure over configuration space with independent positions, so that $\mu_N = \mu_1^N = (4\pi)^{-N}$.
It is useful to write the normalization equation for $\mu_1$ then to introduce spherical coordinates,
\begin{align}
    1 &= \int\limits_{S_2} d^2 \Omega \,  \mu_1(\Omega)\\
    &= \int\limits_{-\pi}^\pi d\phi \int\limits_{0}^\pi d\theta \sin\theta \mu_1(\theta, \phi) \\
    &= \int\limits_{-\pi}^\pi d\phi \int\limits_{-1}^1 du \mu_1(\arccos(u), \phi)
\end{align}
This last expression maps the problem of sphere point picking onto a standard Euclidean point picking problem, and is thus useful to generate Poisson point patterns, using:
\begin{align}
    \phi_i &= 2\pi u_i \\
    \theta_i &= \arccos\left[2 v_i - 1\right]~\label{eq:Poisson_theta}
\end{align}
with $u_i$ and $v_i$ random real numbers drawn uniformly in $[0;1]$.

In particular, note that the DOS of a sphere can be recovered as the distribution of $\theta$ in a Poisson point pattern.
Introducing the cumulative probability distribution function $F(\Theta) = \mathbb{P}[\theta \geq \Theta]$, and inverting the relation between $\theta$ and $v$ in Eq.~\eqref{eq:Poisson_theta},
\begin{align}
    v = \frac{1}{2}\left(1 +\cos\theta \right)
\end{align}
one has
\begin{align}
    F(\Theta) = \mathbb{P}[u \geq \frac{1}{2}(1+\cos\Theta)] = \frac{1}{2}(1-\cos\Theta)
\end{align}
and thus
\begin{align}
    f(\theta) = F'(\theta) = \frac{\sin\theta}{2}.
\end{align}

\section{Spherical Cell List \label{app:CellList}}

\begin{figure}
    \centering
    \includegraphics[width=0.98\columnwidth]{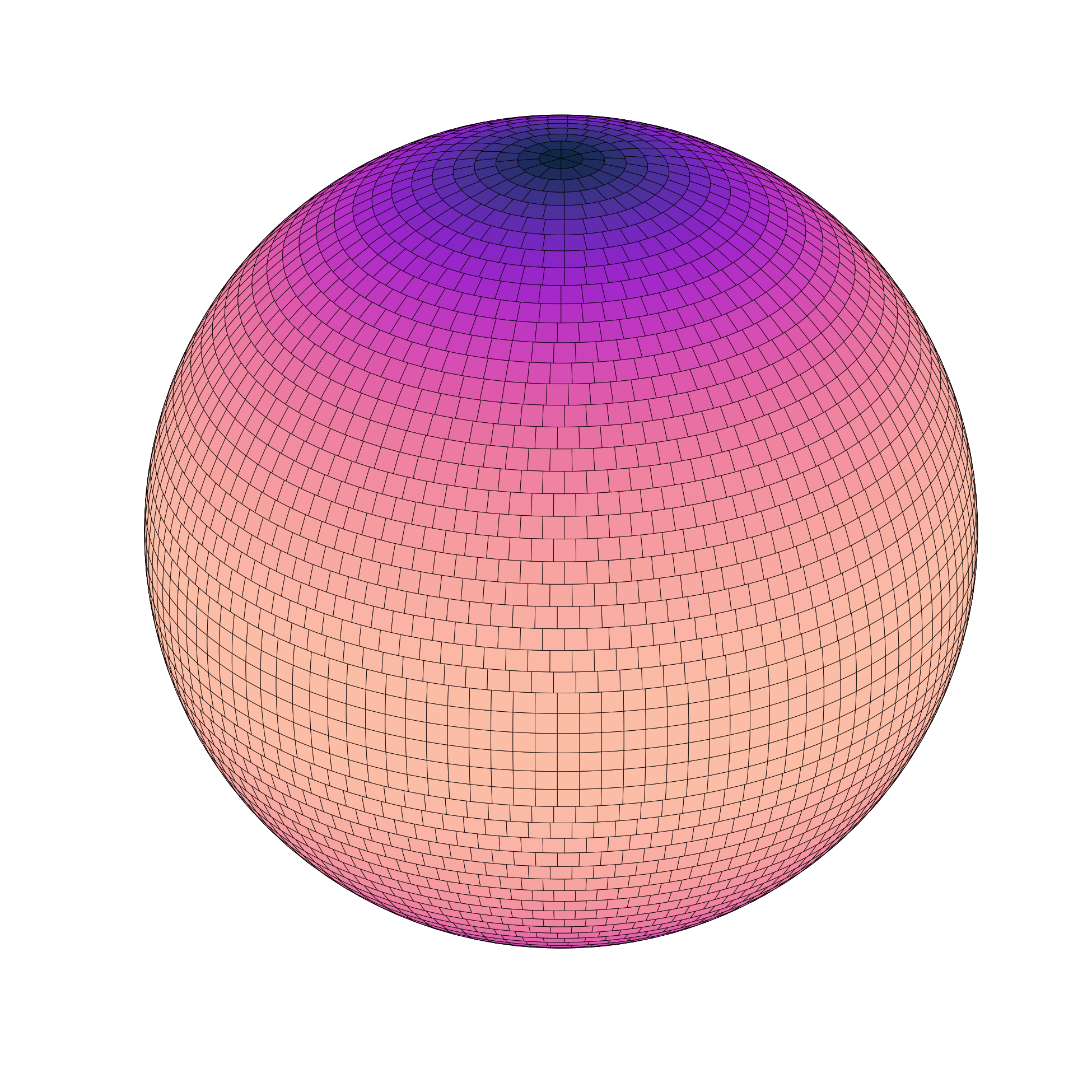}
    \caption{\textbf{Cell list.}
    Example of our choice of spherical cell list.
    We divide the altitude into rings with the arclength corresponding to the largest interaction range.
    Each ring is then subdivided into azimuthal sectors, with a width bounded from below by the largest interaction range.
    In this example, the cutoff is chosen at $0.05$ radians.
    Colors here qualitatively indicate the number of cells along the azimuthal direction.
    }
    \label{fig:CellList}
\end{figure}

In order to evaluate the energy and forces due to finite-range pair potentials in $O(N\log N)$ instead of $O(N^2)$ operations, it is common in flat space to introduce a cell list -- a structure that allows to filter possible neighbors at a small cost that relies on a partition of space.
Here, we adapt this strategy to the surface of the sphere by defining our own cell list.
The idea is the same as in flat space: define cells such that neighbors can lie at most one cell away in every direction.
To achieve this, for a given interaction range $\sigma$, we divide the sphere along the longitude-latitude directions.
Specifically, we first divide the latitude direction into $\left\lfloor\pi/\sigma\right\rfloor$ regular annuli, so that each of them is at least $\sigma$-wide.
Then, we divide the azimuthal direction on each annuli into sectors that are at least $\sigma$-wide on its shortest side.
Concretely, we divide them into $\left\lfloor 2 \pi \sin \theta_{\text{short}} / \sigma \right\rfloor$ equal-length sectors.
The cell list is illustrated in Fig.~\ref{fig:CellList}.
A particle with coordinates $(\theta, \phi)$ is thus mapped unambiguously into a cell with indices $(n_\theta, n_\phi)$, with a small amount of neighboring cells that contain the totality of the possible neighbors of the cell.

\section{Hybrid constraint with Power-Law hyperuniformity and Spherical Percus-Yevick Equation\label{app:PY}}

Following Ref.~\cite{Shih2023}, in order to obtain continuous structure factors with continuous slopes when jointly imposing power-law hyperuniformity and repulsion, we define a constraint
\begin{align}
    S_0(\ell) = B(\ell_{\max}) \begin{cases}
        A\, \ell^\alpha & \text{if }\ell \leq \ell_{PY}, \\
         S_{PY}(\ell) & \text{if } \ell_{PY} \leq \ell \leq \ell_{\max},
    \end{cases}
\end{align}
where $S_{PY}$ is the Percus-Yevick analytical theory for hard disks on the sphere~\cite{Lishchuk2006,Tarjus2012}, $\ell_{PY}$ is the first value of $\ell$ such that the slope of $S_{PY}$ is $\alpha$, $A$ is a factor that ensures that the power law meets the Percus-Yevick value.
The Percus-Yevick theory is computed at the same filling fraction $\eta$ that the repulsive part of the loss uses, so as to produce consistent constraints in real and reciprocal spaces.
Furthermore, we introduce $B$, which is dynamically chosen based on the unconstrained value of the initial $S_{\ell}$ in a small interval $[\ell_{\max}+1; \ell_{\max} + p]$ with $p$ an integer, such that 
\begin{align}
    S_0(\ell_{\max}) = \frac{1}{p} \sum\limits_{\ell = \ell_{\max}+1}^{\ell_{\max}+p} S_{\ell}.
\end{align}
This additional prefactor ensures that the final constrained value is pinned to the value that the unconstrained modes naturally select right outside the constrained window, and thus produces continuous power spectra after optimization.

\section{Spherical Voronoi Tessellation \label{app:Voronoi}}

In order to use farthest-first reinsertion, and to characterize some point patterns, we introduced a spherical Voronoi tessellation.
We note that we here avoided using a $3d$ tessellation, and used a $2d$ one instead, relying on the fact that the stereographic projection is a conformal transformation.
As a result, one may translate spherical coordinates of points on the sphere into their stereographic projection on the plane, establish the Voronoi tessellation there, then back project it onto the sphere.
Due to conformality, the topology of the tessellation is preserved in the operation so that neighborhoods (in terms of point indices) are preserved.
One however needs to be wary of the fact that the stereographic projection sends a pole of the sphere to infinity, thus making the Voronoi tessellation through the stereographic projection generically unstable if one does not heed that fact.
There is however a trick to make the tessellation precise in spite of the pole being sent to infinity: it consists in using one of the points in the pattern as the pole for the projection (a fact noted for instance in a blog post, Ref.~\cite{blog-voronoi}).
Then, the flat Delaunay triangulation of the stereographic projection will have a polygonal hole with $p$ vertices due to the pole sent to infinity, which will contain the point that was placed there.
However, taking advantage of the pole and the point that was excluded from the triangulation are one and the same, it is easy to re-add the point: it is a vertex of all $p$ triangles that connect a side of the polygon forming the hole to the point.
Using the point as the pole thus allows to rigorously reconstruct the whole Delaunay triangulation / Voronoi tessellation on the sphere using a fast $2d$ Delaunay / Voronoi library.
This variant is implemented in our open-source library rusted\_core~\cite{rusted_core}.

\end{document}